\documentclass{WileyMSP-template}
\usepackage{float}
\usepackage{placeins} 
\begin{document}

\pagestyle{fancy}

\title{Sub-40nm Nanogratings Self-Organized in PVP-based Polymer Composite Film by Photoexcitation and Two Sequent Splitting under Femtosecond Laser Irradiation}

\maketitle


\author{Li-Yun Chen$\ddagger$,}
\author{Cheng-Cheng Guo$\ddagger$,}
\author{Ming-Ming Pan,}
\author{Chen Lai,}
\author{Yun-Xia Wang,}
\author{Guo-Cai Liao,}
\author{Zi-Wei Ma,}
\author{Fan-Wei Zhang,}
\author{Jagadeesh Suriyaprakash,}
\author{Lijing Guo,}
\author{Eser Akinoglu,}
\author{Qiang Li$^*$,}
\author{ Li-Jun Wu$^*$}

\begin{affiliations}
  L.-J. Wu, Q. Li, L.-Y. Chen, C.-C. Guo, M.-M. Pan, C. Lai, Y.-X. Wang, G.-C. Liao, Z.-W. Ma, F.-W. Zhang, 
  J. Suriyaprakash\\
  Guangdong Provincial Key Laboratory of Nanophotonic Functional Materials and Devices School of Information and Optoelectronic Science and Engineering\\
  South China Normal University\\ 
  Guangzhou 510006, China\\
  
  Email: ljwu@scnu.edu.cn; liqiangnano@m.scnu.edu.cn\\
 
  \medskip
  L.J. Guo, E. Akinoglu\\
  International Academy of Optoelectronics at Zhaoqing\\
  South China Normal University\\
  Guangzhou 510006, China\\
\end{affiliations}



\begin{abstract}

  Laser-induced periodic surface structures (LIPSSs) on various materials have been extensively investigated because of their wide applications. The combination of different materials allows for greater freedom in 
  tailoring their functions and achieving responses not possible in a homogeneous material. By utilizing a femtosecond (fs) laser to irradiate the Fe-doped Polyvinyl Pyrrolidone (PVP) composite film, highly regular 
  ultrafine nanogratings (U-nanogratings) with a period as small as 35.0 ($\pm$ 2.0) nm can be self-organized on the surface with extremely high efficiency. The period of the U-nanogratings can be controlled by varying 
  the scanning speed of the laser beam (deposited energy) and the thickness of the composite film. Based on the experimental, theoretical, and simulation results, we propose a two-step formation mechanism: composite 
  film excitation and two sequent grating-splitting. The high photosensitivity and low glass transition temperature of the composite film facilitate the fabrication of the ultrafine nanostructures. The proposed design 
  method for the composite material and fabrication process could not only provide a strategy for obtaining highly regular U-nanogratings, but also offer a platform to explore the interaction physics between ultra-short 
  pulses and matter under extreme conditions.

\end{abstract}


\section{Introduction}
Nanostructures and nanodevices have become key components in optics, optoelectronics, plasma science, biochemical detection and biomedicals. Nanogratings are especially attractive because of their widespread applications 
in photonics[1] and biophotonics.[2] They can be applied as typical passive optical elements in high-resolution encoders, spectroscopes, holograms and switching elements.[3,4] They can also be utilized as tools to study 
cell–nanotopography interactions such as adhesion, migration, reorganization, polarization and so on.[5] These applications have fueled the rapid advancement of nanofabrication technologies.[6] Laser assisted processing, 
including laser holography,[7] pulsed laser direct writing[8] and projection,[9] have attracted particular interests due to its low-cost, high flexibility, high efficiency and environment-friendliness. The obtained minimum 
size of a single line can be much less than the irradiation wavelength. For example, it has been reported that a line width as narrow as 1/50 of the wavelength can be realized.[10] However, if they are produced through 
one-by-one multiple scanning without combining two coaxial beams (in the writing systems) with photoinhibitors (in the photoresins), the size of the period is diffraction-limited.[11,12]\\
Laser-induced periodic surface structures (LIPSSs) also known as ripples, have been extensively studied using continuous-wave (CW) or pulsed lasers on a variety of materials, including metals, semiconductors, 
and dielectrics.[13-24] They are categorized into two types in general. One type is termed as low spatial frequency LIPSSs (LSFL) whose period $ \Lambda  $ approximately equal to the laser wavelength and has been attributed 
to the interference of incident light and the scattered light.[17,20] The other one is high spatial frequency LIPSSs (HSFL), which has a period $ \Lambda  $ of less than 1/3 wavelength and are due to the surface waves induced 
by laser-excited electrons.[19,21,22] When ultra-short laser pulses such as femtosecond (fs) are applied, the size of LIPSSs can be much smaller than the wavelength being irradiated. In contrast to the plethora of
 knowledge on laser processing of other materials in the fs regime, there are only a few works on polymer processing.[23,26] \\
In LIPSSs such as nanogratings, when the power of irradiation is close to the materials’ ablation threshold, the energy is prone to be concentrated on the ridges, causing splitting of the initially formed gratings. 
This is a common phenomenon. Whereas it can serve as a possible way for fast reducing the period of the gratings, normally only one complete splitting can be observed. The secondary splitting creating complete and 
regular nanogratings bears a low probability due to the thermal effect.[27] To the best of our knowledge, there is no study has yet reported the splitting phenomenon in polymers.\\
On the other hand, the combination of different materials allows tailoring their functions with greater freedom, which can achieve responses not achievable in a homogeneous material. Polymers or glasses are commonly 
used as the matrix for their supporting and scaffold abilities.[28-29] In order to obtain more functions, a polymer matrix owning a high miscibility with different solutions are highly desirable. Polyvinyl Pyrrolidone 
(PVP) is a biodegradable and water-soluble polymer. It is widely used in medicine, cosmetics and the food sector. In addition to being hydrophilic, PVP has excellent solubility in solvents with different polarities 
and good binding properties.[30] Therefore, it is very easy to be blended with various aqueous solutions, whereby rich functions can be obtained in the PVP-based polymer composites.[31] For example, PVP has been mixed 
with Pt/Pd solutions for site-specific catalysis after being irradiated by a pulsed laser.[32] It has also been blended with noble metallic aqueous solutions such as Au and Ag followed by pulsed lasers irradiation 
process to obtain metallic nanostructures.[33-36] The interaction between the materials and laser is a photoreduction process through which the metallic ions can be reduced into atomic states and metallic nanoparticles 
can be formed finally. PVP functions as a supportive or capping material in these cases. However, PVP itself has an absorption spectrum in the region of 200-280nm and can be photo-crosslinked upon irradiation of UV 
light.[37] By introducing Fe or Cu ions within, its photosensitivity in the visible spectrum region can be enhanced.[38] \\
In this paper, we investigate the interaction between the PVP-based composite film and fs pulsed lasers with two different repetition rates, in which different spatial profile of the focused Gaussian beam is provided 
to explore the mechanism behind the nanogratings. The addition of Fe ions into the PVP polymer not only increases its photosensitivity[38] but also lowers its glass transition temperature Tg, which reduces the bonding 
energy of the solidified PVP composite. Furthermore, the bonding mode of the coordination bond between Fe ions and PVP provides a bridge between the fs laser pulses and the composite for ultrafast subcycle energy 
transfer, which greatly reduces the thermal effect. Highly regular U-nanogratings with a period as small as 35.0 ($\pm$ 2.0) nm can be formed in the composite film by controlling the irradiation fluence to approach its 
ablation threshold. The period is less than 1/20 of the irradiation wavelength and is the smallest value reported. The orientation is always perpendicular to the polarization direction of the irradiation beam. The 
period of the U-nanogratings can be tuned by adjusting the scanning speed of the beam and the thickness of the film. According to the experimental results, theoretical analysis and simulation results, we propose 
that the highly excited electrons which induce the excitation of surface plasmons (SPs) are responsible for the formation of HSFL, and the subsequent two splitting can reduce the period very fast. These two processes 
provide a complete picture of the U-nanogratings formation mechanism. The material and fabrication design method presented here could not only provide a strategy for obtaining highly regular U-nanogratings, but also offer 
a platform to explore the interaction physics between ultra-short pulses and matter under extreme conditions. \\

\section{Experimental Section}
Processing of Materials: The composite film is composed of PVP and Fe(NO3)3·9H2O  with a weight ratio of 1:1. PVP$( $Shanghai EKEAR$) $and Fe(NO3)3·9H2O (Shanghai Crystal Pure) were mixed in deionized water and stirred for 20 
mins. Then they were spin-coated on the cover glasses and dried at 100°C for 10 mins to form the composite film. After irradiated by the fs laser, the composite film was soaked in ethyl alcohol for two hours and the 
unsolidified region was then washed out by deionized water. The thickness of the film was controlled by adjusting the concentration of PVP and the spinning speed.\\
Characterization of the Materials and Nanostructures: The thickness of the film and the height of the nanostructures were measured by an atomic force microscopy (AFM, JPK nanoscribe) at a tapping mode. The surface 
morphology of the nanostructures was examined by a scanning electron microscope (SEM) (Ultra 55, Zeiss) and AFM. The static refractive index of the composite film was measured by an ellipsometer (L116S300 STOKES) 
at a wavelength of 632.8 nm. As 632.8 nm is close to 800 nm and there is no abrupt change for the static refractive index in this wavelength range, we utilize the measured value at 632.8 nm in the paper.\\
Laser Irradiation Experiments: We chose two fs laser sources for irradiation. The one with 76 MHz (at wavelengths of 532, 610 and 800 nm respectively) repetition rate was utilized to scan the sample in order to 
ensure a uniform laser irradiation distribution along the scanning line, which was realized by translating the substrate in the plane parallel to the sample surface. Figure S1a plots the schematic of the writing
 system. Laser pulses (130 fs) were delivered from a mode-locked Ti: sapphire oscillator (800 nm, Mira-HP, Coherent) and a tunable OPO (505-740 nm, Mira-OPO-X). An attenuator and a half-wave plate were used to 
 control the energy and the polarization direction of the laser beam. The polarized laser beam was focused normally on the sample by a 100x oil-immersion objective (NA=1.4, Zeiss). The diameter D of the focused 
 spot was about 1 µm $(1/e^2)$. Based on the laser power P measured before entering into the microscope, the laser fluence can be estimated by $F=P/(\pi \times (D/2)^2\times 76\times 10^6)$. The sample was fixed on a computer-controlled 
 three-dimensional high-precision nanopositioning translation stage (P-563, PI). The other one with a 1 kHz repetition rate was applied to investigating the formation mechanism of the U-nanogratings at a single 
 spot irradiation mode. As shown in Figure S1b, laser pulses at a central wavelength of 800 nm with a repetition rate of 1 kHz (90 fs) were delivered from a Ti: sapphire amplifier (Legend, Coherent). The energy 
 and the polarization direction of the laser beam were also controlled by an attenuator and a half-wave plate. The polarized laser beam was focused normally on the sample by a 125 mm focal length lens. The diameter
  of the focused spot was about 25 µm (1/e$^{2}$). All the irradiation experiments were carried out in air at room temperature.\\
Numerical simulations: The numerical simulations of the field distribution under 800 nm irradiation were implemented by commercially available software, finite difference time domain (FDTD) solutions (Lumerical). 
The geometric parameters of the model were chosen according to the AFM and SEM results. The dielectric constant was derived from the Drude model.\\
\section{Results and discussion} 
\FloatBarrier
\subsection{Formation of the U-nanogratings by the fs laser with a repetition rate of 76 MHz}
Figure 1 exhibits seven typical SEM images for the lines written on the composite film under different conditions. The irradiation wavelength was 800 nm 
in all the cases if not stated otherwise. When the laser fluence reaches the threshold of solidification (16 $mJ/cm^2$ at a speed of  4 $\mu m/s$), 
the composite film was solidified and cannot be solved by its solvent (either water or alcohol), leaving a line as shown in Figure 1a. 
By increasing the laser fluence to 25 $mJ/cm^2$, some nanoholes were formed in the center of the solidified line marked by the green 
dashed circles (Figure 1b). Further increasing the fluence to 75 $mJ/cm^2$ gives rise to extremely regular U-nanograting as can be observed
 in Figure 1c. The period $\Lambda $ of these U-nanogratings reaches as small as 43$\pm $4 $nm$ (the measuring method is described in Figure S2) in this
  figure. This period is far beyond the optical diffraction limit. Furthermore, as multiple periods can be formed in one scanning, t
  he fabrication efficiency of the nanogratings is extremely high. When the fluence was increased to 100 $mJ/cm^2$, the U-nanogratings 
  were found to be ablated (Figure 1d). The regular nanogratings always appeared in the center of the laser irradiation track in our 
  experiments, indicating that the required energy is higher than that for solidification. The direction of the periods is definitely perpendicular 
  to the polarization[39] as shown in Figure 1c-1f. With a circularly polarized irradiation beam, the U-nanogratings broke into small nanosquares and 
  is in agreement with previous reports (Figure 1g).${[40-41]}$\\
\begin{figure}[htp]
  \includegraphics[width=0.8\linewidth]{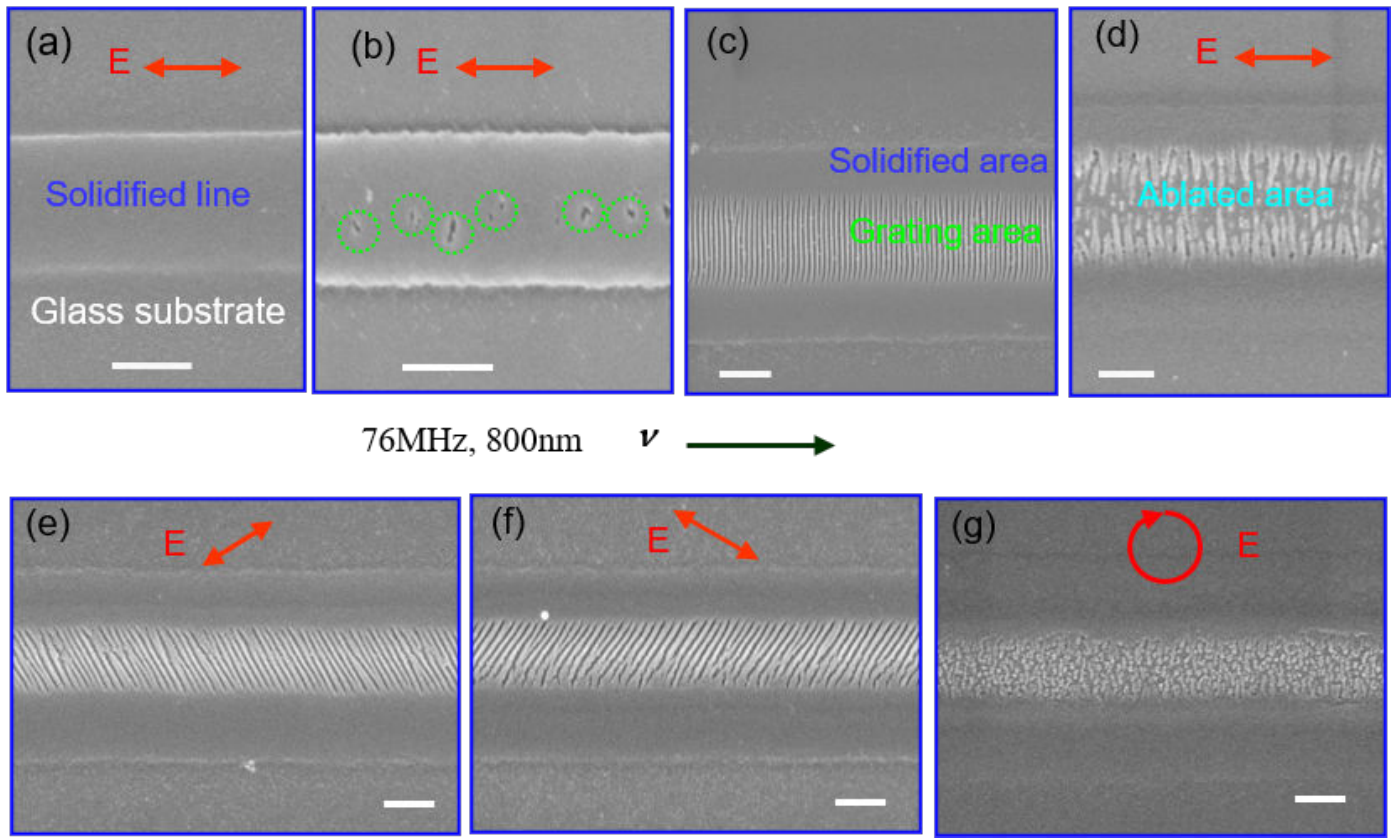}
  \centering
  \caption{SEM images of the lines written by scanning the sample with a speed of 4 $\mu m/s$ when the fluence was (a) 16 $mJ/cm^2$, (b) 25 $mJ/cm^2$, 
  (c) 75 $mJ/cm^2$ and (d) 100 $mJ/cm^{2}$. (e-g) demonstrate the influence of the polarization direction on the formation of the 
  nanogratings/nanostructures. All the two-way red arrows point to the polarization direction of the laser beam in the paper. The scale 
  bars in the SEM images represent 400 nm. The thickness of the film was fixed at 80 nm unless otherwise specified.}
  \label{fig:boat1}
\end{figure}
To determine the threshold for producing U-nanogratings, we correlate one of the SEM results with the spatial intensity profile of 
the focused Gaussian beam in an area with a diameter of ~1 $\mu $m as shown in Figure 2a. The detailed analysis process can be found in 
the Supporting Information. Figure 2b demonstrates that the threshold F* for producing U-nanogratings increases from 38 to 54 $mJ/cm^2$ 
when the scanning speed v increases from 1 to 16  $\mu m/s$. The upward trend is expected since a faster scanning speed means fewer pulses 
accumulated on the materials, thus higher fluence is required.\\
\begin{figure}[H]
  \includegraphics[width=0.9\linewidth]{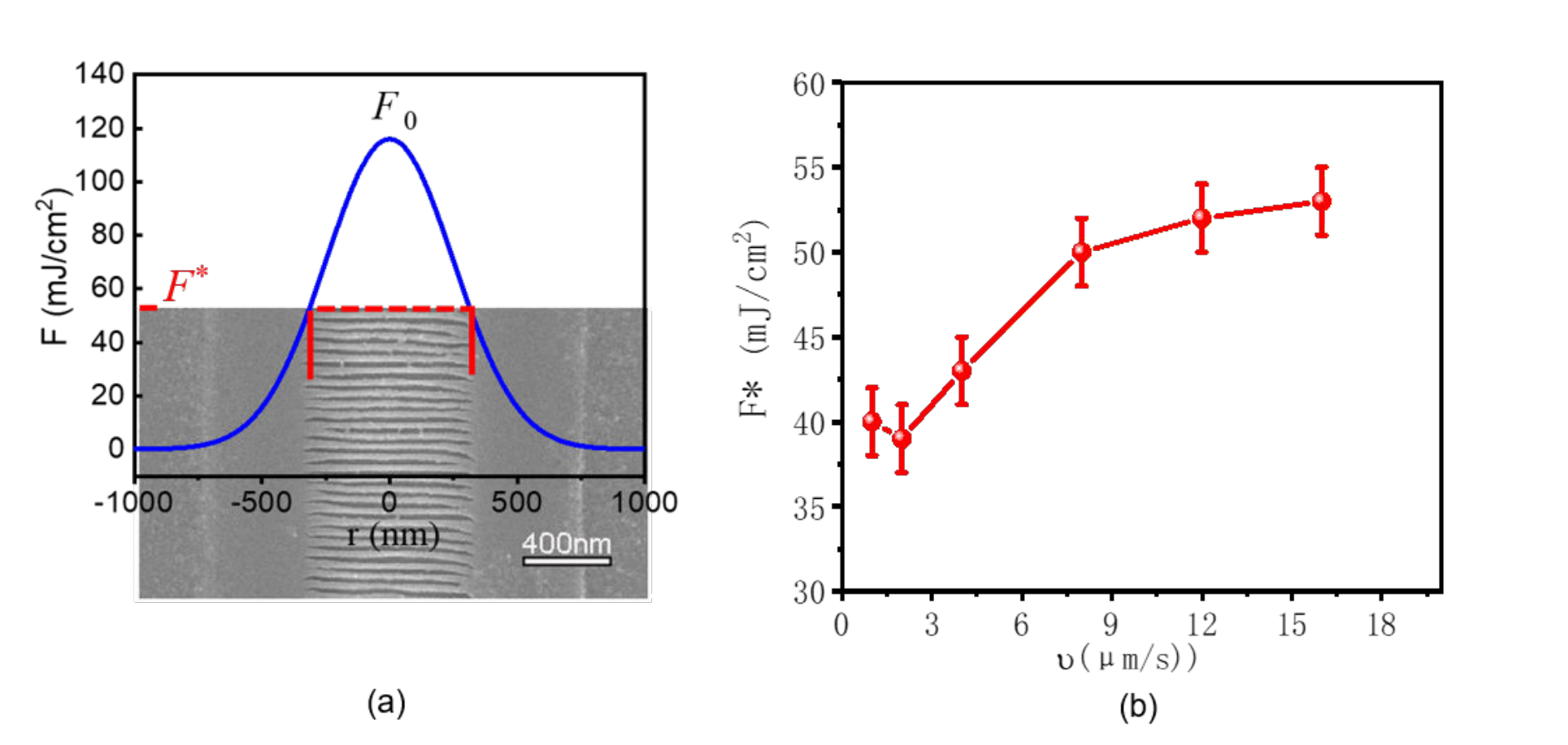}
  \centering
  \caption{(a) Calculation of the threshold fluence (F*) based on a Gaussian profile (blue line) and the SEM result. 
  (b) The threshold of forming the U-nanogratings as a function of the scanning speed v when the average fluence is $67 mJ/cm^2$.}
  \label{fig:boat1}
\end{figure}
The irradiation wavelength, scanning speed v, irradiation fluence F, and thickness of the spin-coated film H were then modified to 
investigate the critical parameters controlling the period and quality of the nanogratings. Figure 3a summarizes the obtained periods 
under different experiment conditions as a function of v. It can be observed that $\Lambda $ is decreased with an increase of v when irradiated 
by different wavelengths under different fluences. The period $\Lambda $ increases as the thickness H of the composite film increases (Figure 3b).
 The corresponding uniformity of the formed nanogratings varies as shown in Figure S3. When H is smaller than 60 nm, only irregular 
 nanostructures can be formed (Figure S3a). While if the film is excessively thick, such as H=300 nm, the center of the line was 
 entirely ablated, as shown in Figure S3f. The height h of the nanogratings is also shown in Figure 3b (right axis) and Figure S3. 
 The variation of h is only 3.5 nm when H is increased from 80 nm to 230 nm, indicating that the thickness of the composite film does 
 not play an important role here. In Figure 3c, the value of h is around 40 nm and its relation with v, H and F is random. The bottom 
 panel of Figure S3 demonstrates some typical AFM images of the patterns obtained under a series of scanning speeds. The height 
 information for the nanogratings area (marked by the blue dashed line) shows no discernible difference for various v and H, 
 which is consistent with the results displayed in Figure 3c.\\
\begin{figure}[htp]
  \includegraphics[width=\linewidth]{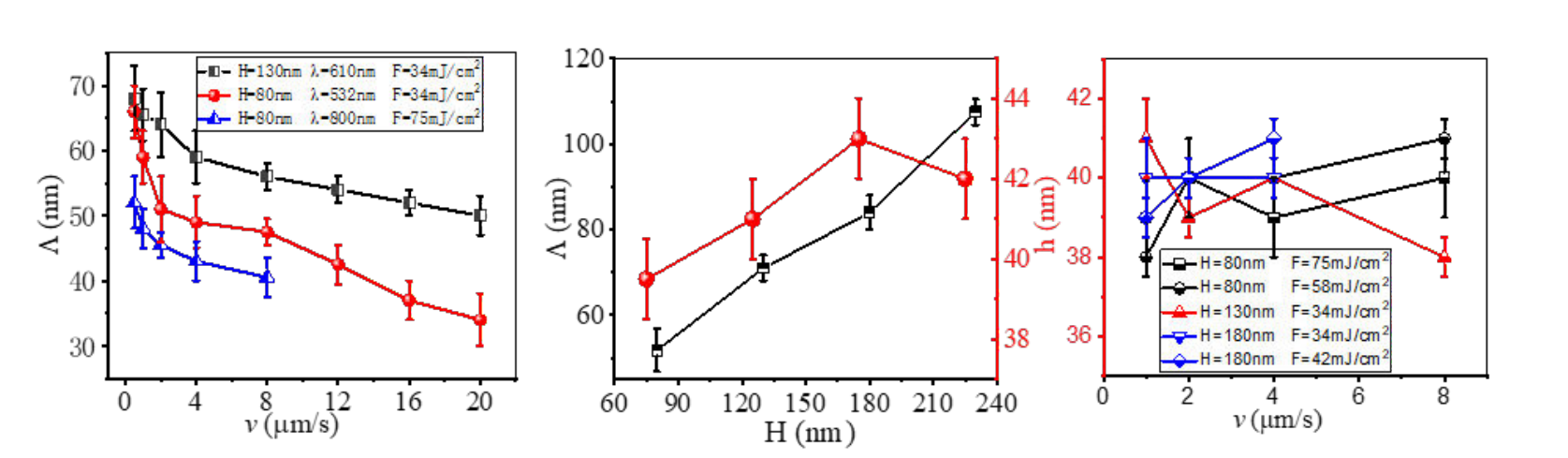}
  \caption{The relationship between the grating period $\Lambda $ and other parameters. (a) The influence of the scanning speed v under 
  different irradiation wavelengths $\lambda  $ and film thickness H. (b) The left axis shows the influence of H (before irradiation) on
  $\Lambda $ and the right one on the height h of the formed U-nanogratings. (c) The height h of the formed U-nanogratings as a 
  function of thescanning speed v under different H and irradiation fluence F.}
  \label{fig:boat1}
\end{figure}
\FloatBarrier
\subsection{Formation of the gratings by the fs laser with a repetition rate of 1 kHz}
Figure 1 demonstrates different stages of the material change, including solidification (Figure 1a), formation of defects (Figure 1b), 
formation of U-nanogratings (Figure 1c), ablation (Figure 1d) upon different irradiation fluences. However, the specific formation 
process of the U-nanogratings cannot be found either by tuning v or F when the repetition rate of the irradiation laser was 76 MHz. 
Taking into account that the diameter of the focused spot for our 76 MHz writing system is only about 1 $\mu $m, it is possible that the 
intensity change rate along the Gaussian spatial profile is too fast, preventing us from monitoring the precise development process 
of the nanogratings. A fs laser source with 1 kHz repetition rate has a much higher peak fluence and the threshold of forming the 
nanogratings can be easily achieved even with a significantly larger focused spot. Figure 4 provides a comparative analysis of the 
spatial profile of a focused Gaussian beam with two different repetition rates. Obviously, at 1 kHz, the change rate of the intensity 
distribution (from spatial) is much slower.\\
\begin{figure}[H]
  \includegraphics[width=0.6\textwidth]{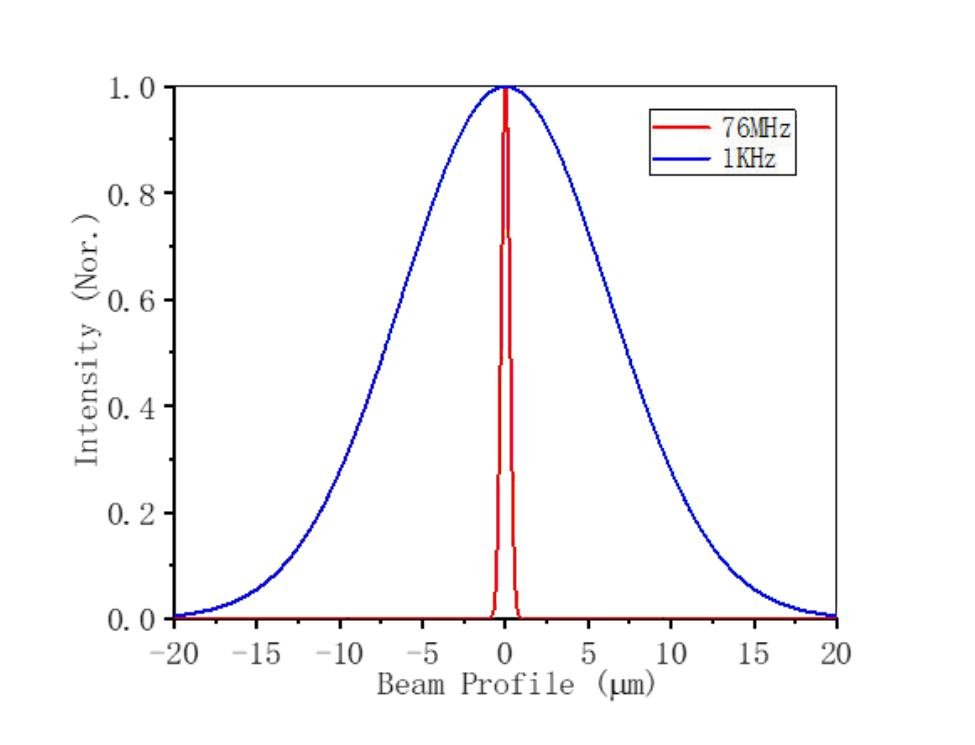}
  \centering
  \caption{The Gaussian beam spatial intensity profiles of the focused spot under the repetition rates of 76 MHz and 1 kHz. The 
  intensity is normalized to the peak value.}
  \label{fig:boat1}
\end{figure}
Figure 5a-c display the morphology of the composite film irradiated by a fs stationary laser beam at a repetition rate of 1 kHz with 
different pulse numbers. The initial thickness of the film and the fluence were 70 nm and 780 $mJ/cm^2$, respectively. For the single 
spot irradiation experiment, the pulse number N was used to evaluate the energy deposited onto the film and can be tuned by the computer.
 The defect-holes start to emerge at the center of the solidified area at a relatively early stage of laser irradiation, such as N=1000 
 (Figure 5a). At the brim area in which the light intensity is lower, the defect-holes are distributed randomly. From the periphery to 
 the center, we can observe that the defect-holes are gradually lined up into order as marked by the orange arrows and evolve into 
 nanogratings at the center of the irradiation spot. Similar results have been reported in the bulk glasses.[42] We name them as 
 “coarse” nanogratings (abbreviated as C-nanogratings) as their periods are much larger than the above U-nanogratings. When N is 
 increased to 5000 as shown in Figure 5b, the periphery of irradiated area is covered with C-nanogratings and U-nanogratings appear 
 in the center. With a further increase of the pulse number (N=60000), those initially formed C-nanogratings ($\Lambda _{1}$=240 nm) 
 start to be split. The splitting can be self-replicated if the energy is high enough. The first and second splitting processes are 
 clearly observed from the edge to the center of the irradiated spot as shown in Figure 5c1. \\
 \begin{figure}[H]
  \includegraphics[width=0.8\textwidth]{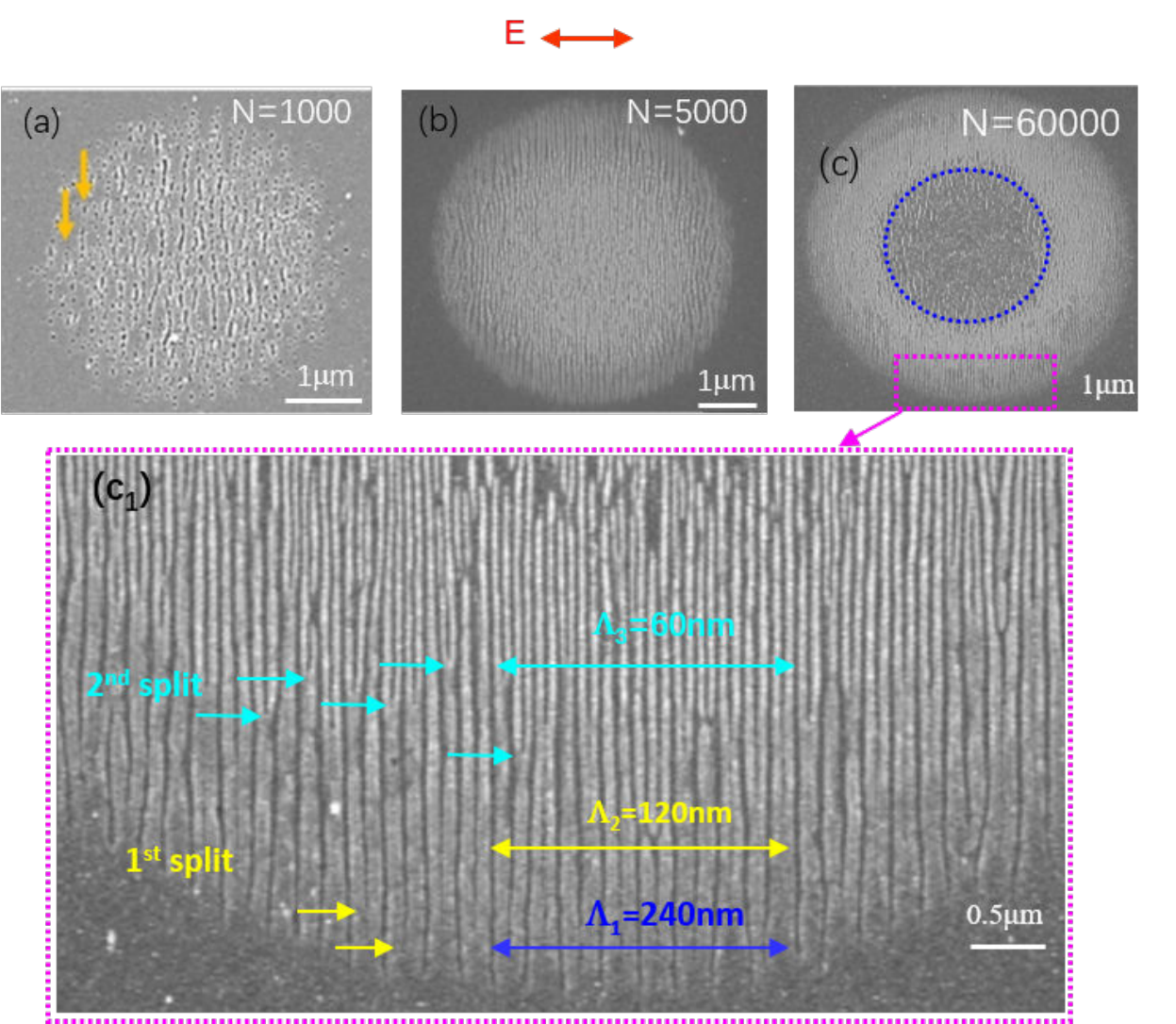}
  \centering
  \caption{SEM images of the single spots irradiated by the fs laser at a repetition rate of 1 kHz with different pulse numbers N. 
  The fluence was 780 $mJ/cm^2$. (a) N=1000. (b) N=5000. (c) N=60000. The blue dashed circle surrounds the ablated area. With an increase 
  of N, hole-defects, nanogratings and ablation appear consequently. (c1) illustrates the details of the periphery of the grating-area. 
  Yellow/cyan arrows mark some of the 1$^{st}$/2$^{nd}$ splitting positions.}
  \label{fig:boat1}
\end{figure}
Although the first split has been observed in many material systems, including semiconductors and metals,[27, 43]  the second split 
has mostly been found in the transparent glasses upon fs laser irradiation.[44] Due to the self-replication of splitting, the period 
of the C-nanogratings can be reduced very quickly to be 60 nm. This value is in a similar order to the period obtained by the 76 MHz 
fs laser. Since the laser intensity increases from the periphery to the center of the irradiation spot, the appearance of splitting 
should be related with the deposited energy. However, as seen by the cyan arrows in Figure 5c1, the place to discover the same-level 
split does not strictly follow the energy trajectory of the Gaussian beam. This argues that the initiation of the splitting is not 
only determined by the energy but also closely related to the film’s state. If the deposited energy is excessively large, such as at 
the center (enclosed by a dotted blue circle) of the irradiation spot shown in Figure 5c, the film is to be ablated completely.\\
\begin{figure}[H]
  \includegraphics[width=0.85\textwidth]{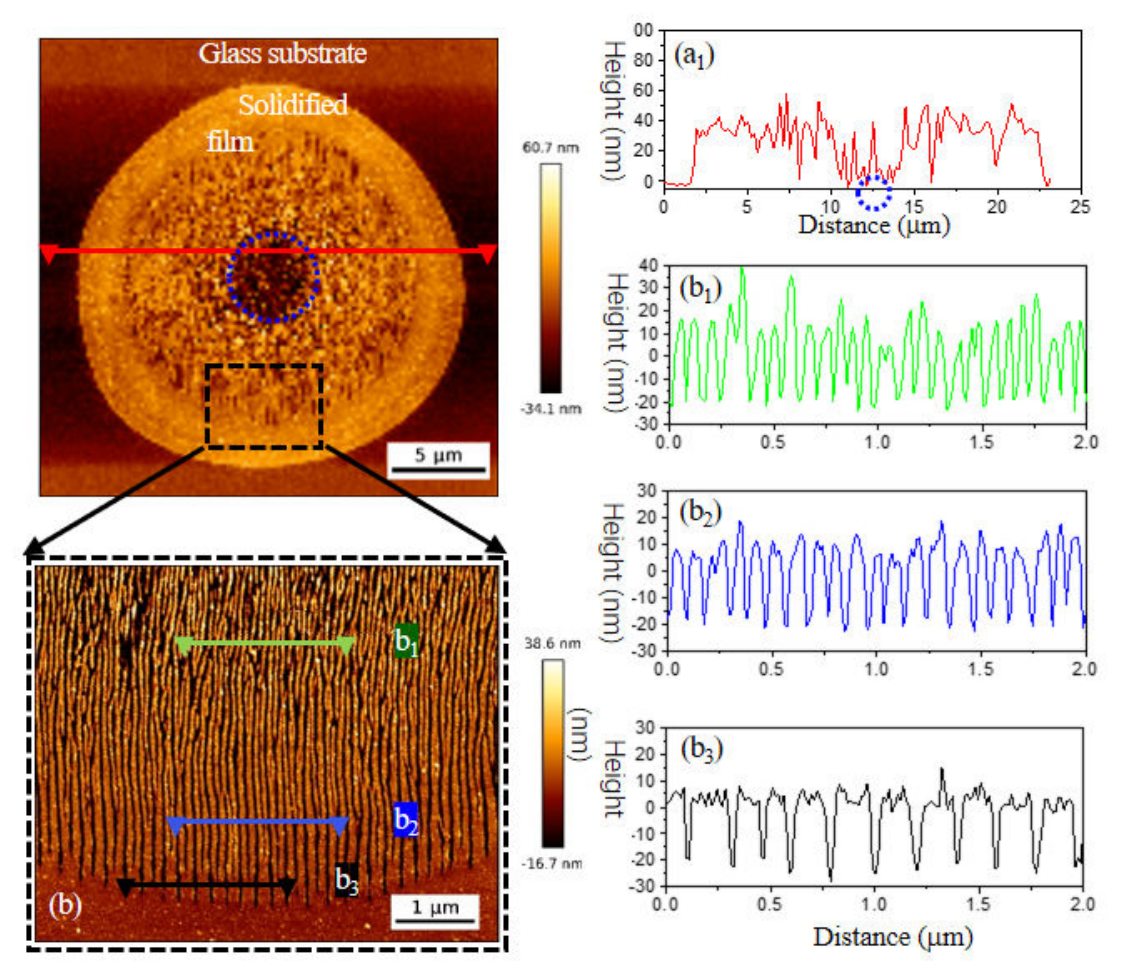}
  \centering
  \caption{AFM images of a typical spot after irradiation at a fluence of 600 $mJ/cm^2$ and N=100000. (b) shows the enlarged 
  view of the periphery of the grating-area. (a$_1$), (b$_1$) and (b$_2$) plot the height along the corresponding lines marked in 
  (a) and (b) respectively. }
  \label{fig:boat1}
\end{figure}
Figure 6a shows the AFM image of a typically irradiated spot. An enlarged view of the periphery of the grating-area is shown in 
Figure 6b. The curves with different color on the right panel plot the height information along the same color lines marked on the 
left figures. The red line describes the coarse scanning (Figure 6a1). It shows that the thickness of the solidified film is $\sim$40 nm, 
which is similar to the height of the U-nanogratings shown in Figure S3. The film has almost utterly ablated at the center of the 
irradiation spot (marked by a blue dashed circle). Away from the periphery to the center (marked by the black, green and blue lines in 
sequence), the depth of the nano-grooves is increased from $\sim$25 nm to $\sim$30 nm and $\sim$40 nm, respectively. Based on the method to derive the 
threshold for forming the nanogratings (F*) (refer to Eqs.1-5 in the Supporting Information), we can obtain the threshold of five 
processes: solidification ($T_{0}$), formation of C-nanogratings ($T_{1}$), 1st splitting ($T_{2}$), 2nd splitting ($T_{3}$) and ablation ($T_{4}$), as shown in 
Figure S4.  Figure 7 summarizes the relationship between the threshold fluence and pulses for the five processes. As expected, less 
pulses are required to obtain the same process under higher fluence. The splitting phenomenon does not appear until N is increased to 
be $\sim 1000$.\\
\begin{figure}[H]
  \includegraphics[width=0.6\textwidth]{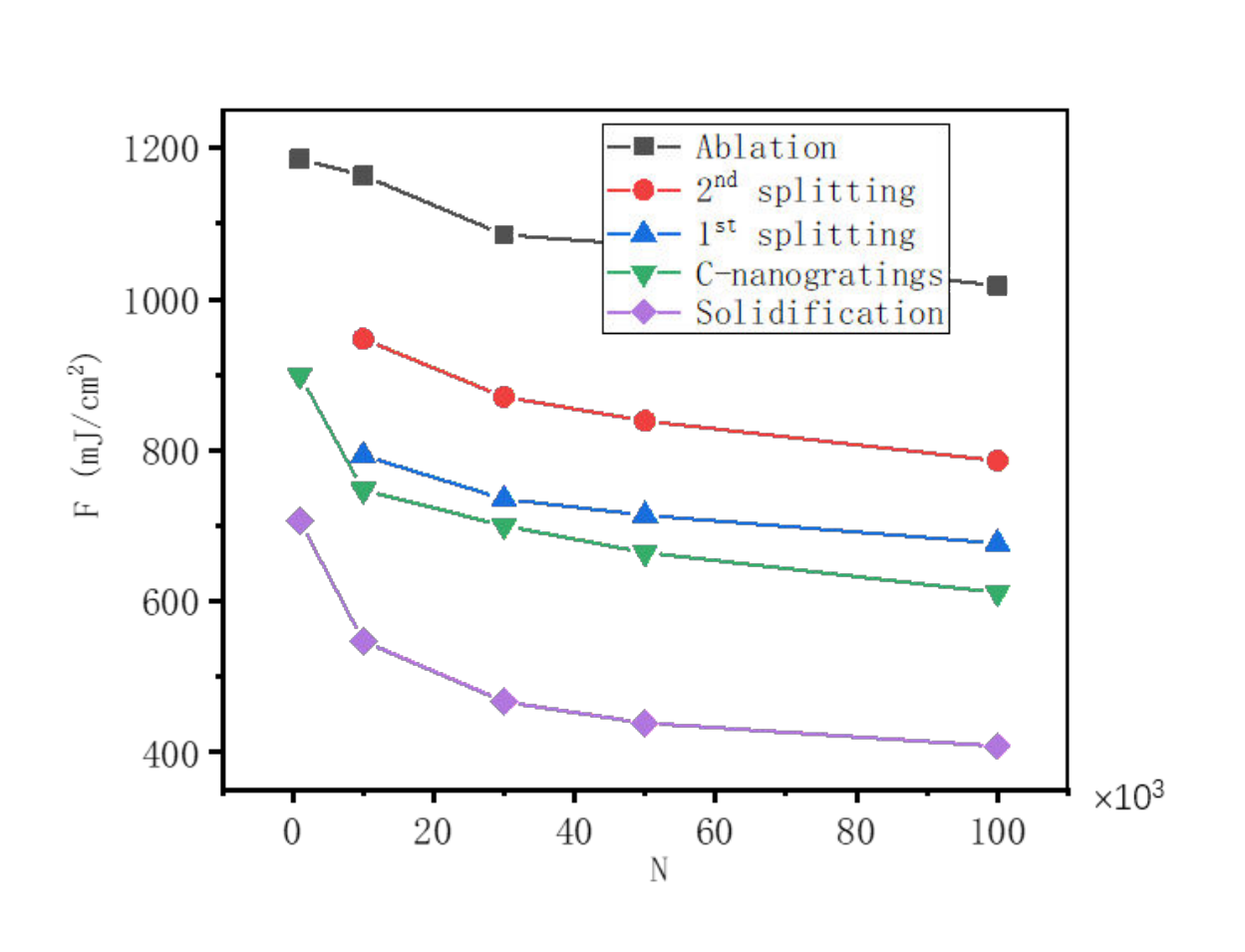}
  \centering
  \caption{The relationship between the required fluence (F) and the number of pulses (N) for
  solidification, formation of C-nanogratings, 1st splitting, 2nd splitting and ablation. }
  \label{fig:boat1}
\end{figure}

\FloatBarrier
\subsection{The formation mechanism of the C-nanogratings}
According to the experimental results obtained by the fs laser with 76 MHz and 1 kHz repetition
rates, we propose a possible forming process of the U-nanogratings in the following context.\\

Our previous paper has reported that the Fe-doped composite polymer film is amorphous after irradiated by a laser beam.[38] For 
obtaining LIPSS in amorphous polymers, the temperature should be higher than its glass transition temperature (Tg) to allow polymer 
chains to rearrange.[45] Figure 8 shows that when Fe ions are doped into the PVP polymer, Tg can be reduced from 202.25°C to 130.28°C, 
which can reduce the required energy for forming LIPSS.[24]\\
\begin{figure}[H]
  \includegraphics[width=0.6\textwidth]{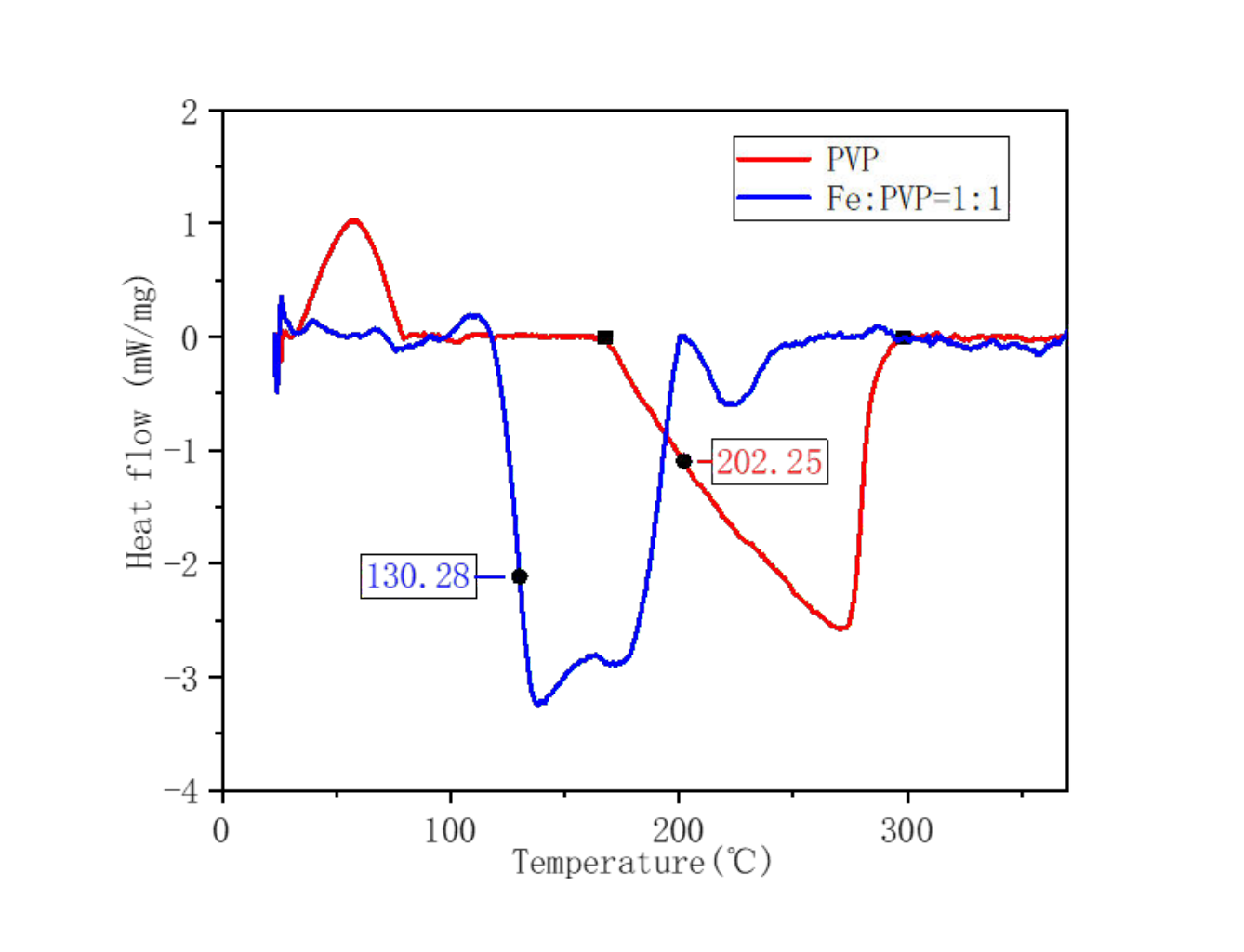}
  \centering
  \caption{Differential scanning calorimetry (DSC) curves of PVP (red) and composite (blue).}
  \label{fig:boat1}
\end{figure}

Although the photonic band gap Eg of PVP is shrunk from 5.0 eV to 2.7 eV when Fe ions are introduced,[38] the Fe-doped composite film 
is initially transparent to the laser pulses and its conduction band is empty. However, if the intensity is sufficiently high, electrons
 in the valence band can be transferred into the conduction band via nonlinear ionization processes under fs laser irradiation.[46,47] Therefore,
  the composite film can be excited into a state similar to that of metals transiently and becomes highly absorptive. The 
 abundant electrons generated by nonlinear processes including multiphoton and avalanche ionization respond collectively to incident 
 light by oscillating in resonance with the electric field of the laser beam, resulting in SP excitation.[48-49] The permittivity of the
  excited composite film can be described by the Drude equation[19,50] and its dielectric constant $\varepsilon^*$ varies with the laser
   excited electron concentration Ne.[22,44,51] \\
Based on Eq. (6)-(8) in the Supporting Information and corresponding analyses, Ne is calculated to be $4\times 10^{21}$ $cm^{-3}$when the material is 
optically broken down and turns from a dielectric to a metallic state. The electron damping time $\tau c$ and the effective electron mass m* 
are set to be 2.5 fs and 0.45m$_e$ respectively.\\
Taking into account that the composite film is much thinner than the irradiation wavelength, light experiences a kind of effective 
refractive index which is determined not only by the thin film, but also by the above air and the below substrate. Its effective 
dielectric constant $\varepsilon$ can be given by,[51] \\
\begin{equation}
  \frac{1}{\varepsilon} =\frac{1}{\varepsilon^{*}}+\frac{1}{\varepsilon_{s}}
\end{equation}
when SPs are excited. $\varepsilon_s$ stands for the effective dielectric constant of the external environment defined by the volume 
fraction of air and substrate. If LIPSS have been formed in the film, the SPs would transport along the interface between air and 
LIPSS.[21] $\varepsilon_s$ can be obtained by,[52] \\
\begin{equation}
  \varepsilon_{s} =a\varepsilon_{1} +(1-a)\varepsilon_{2} 
\end{equation}
where a describes the influence of the substrate. $\varepsilon_{1}$=1 and $\varepsilon_{2}$=2.28 represent the dielectric constant of 
air and the glass substrate. Our experimental results indicate that the height of the nanogratings is as small as $\sim$40 nm. In this 
case, the substrate influence the SPs dominantly and a can be considered as 0. If the film is very thick, the effect of the substrate
 can be ignored thus a can be assumed to be 1. The wavelength of the SP $\lambda_{sp}$ is given by,[51]\\
\begin{equation}
  \lambda _{sp} =\frac{2\pi}{k _{sp}} 
\end{equation}
\begin{equation}
  k_{sp} =\frac{w}{cRe(\sqrt{\varepsilon})} 
\end{equation}

The period of LIPSS $\Lambda $=$\lambda sp$/2 when the wavevector matching condition of the standing plasmon wave along the surface is 
satisfied. The intensity distribution of this standing wave can be imprinted on the film surface if the irradiation power density 
approaches the ablation threshold. \\

 \begin{figure}[H]
  \includegraphics[width=0.7\textwidth]{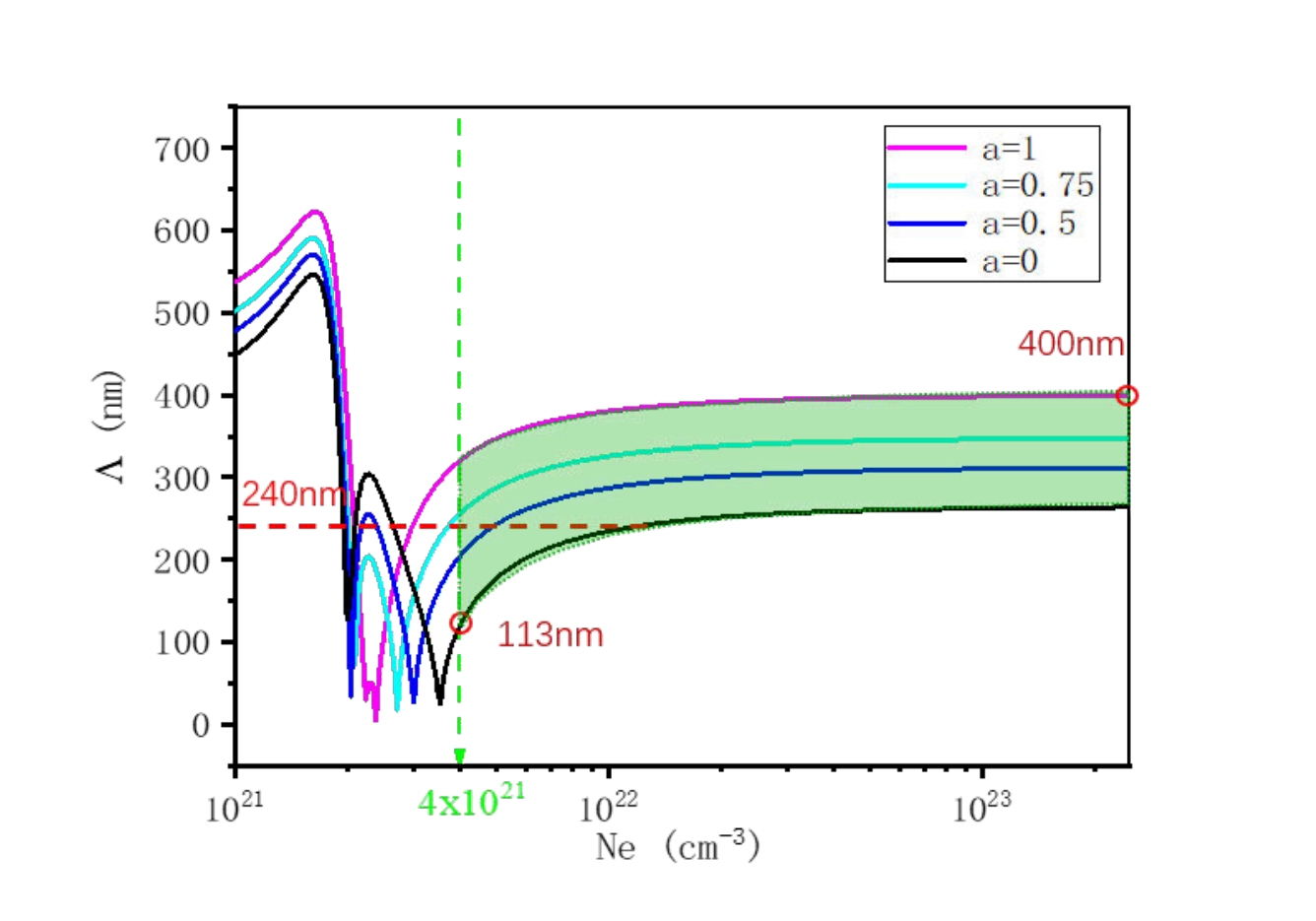}
  \centering
  \caption{The period $\Lambda$ of the initially formed C-nanogratings as a function of the free electron density (Ne) under different volume 
  fractions of the substrate. $\tau_c$=2.5 fs and $m^*$=$0.45me$. The green dashed line marks the required Ne for exciting the composite 
  film. The green shaded region plots a part range of possible $\Lambda$ under our experimental conditions. }
  \label{fig:boat1}
\end{figure}

According to Eqs. (1)-(4) and the above discussions, we plot the relationship between $\Lambda$ of the initially formed C-nanogratings 
and Ne with different a in Figure 9. On account of that, the electrons are excited from the valence band by intense laser pulses, the 
maximum density should be less than that of valance electrons ($2.47$$\times $$10^{23}$ $cm^{-3}$). The obtained period (Figure 5c) 
$\Lambda_{1}$=240 nm is marked as a red dashed line in the figure. It crosses the area with a small a, which is consistent with 
the thin feature of the nanogratings shown in Figure 6. Therefore, the above SP model can explain the period at the periphery of the irradiation spot (refer 
 to Figure 5c). \\
\FloatBarrier
\subsection{Two sequent splitting in the C-nanogratings}
Combining the derived parameters from Eqs.1-7 and the height morphology information from AFM results (Figure 6b), we set up a numerical 
simulation model as shown in Figure 10a to investigate the evolution of the light intensity distributions under continuous irradiation 
(800 nm). At first, only TM excitation being able to induce the energy to be distributed periodically explains why the nanogratings are 
always perpendicular to the polarization of the irradiation laser beam in the experiments. The C-nanogratings with $\Lambda_{1}$=240 nm 
were initially formed and the energy is concentrated in the nano-grooves, as shown in Figure 10b. However, with an increase of h of the 
initially-formed C-nanogratings (Figure 10c), besides being localized inside the original nano-grooves, the energy starts to be 
concentrated on the ridges of the gratings. Two energy maxima appear between the original three nano-grooves. Under this situation, 
new nano-grooves tend to be formed at positions with enhanced intensity, as indicated in Figure 10d. The energy is inclined to be more 
concentrated into the newly-formed shallow nano-grooves, revealing that the growth of the newly-formed nano-grooves is a positive 
feedback until their depth becomes the same as those of the original ones. Therefore, new energy maxima would appear again between the 
nano-grooves with a same depth whereby the splitting process can be self-copied as illustrated in Figure 10e-10g. The period can be 
scaled down quickly to $\sim$60 nm after the initially formed C-nanogratings being split twice. \\
\begin{figure}[H]
  \includegraphics[width=0.95\linewidth]{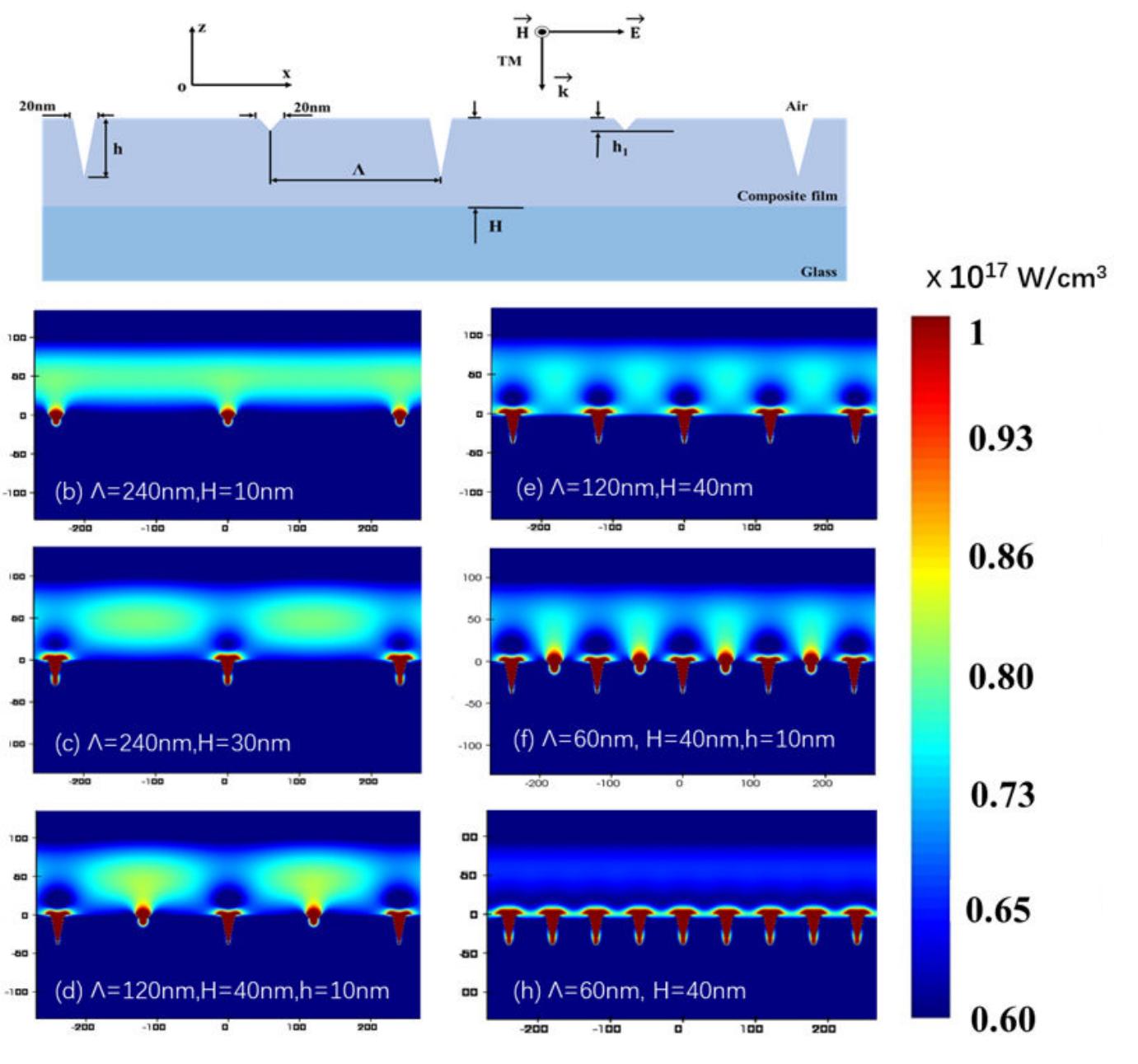}
  \centering
  \caption{The distributions of the light intensity with different parameters under continuous irradiation
  of the TM-polarized plane wave. (a) The definition of the geometric parameters for the nanogratings. (b)
  The intensity is concentrated at the initially-formed nano-grooves while uniformly distributed along with
  the interface between the film and air. (c) With an increase of the depth of the nano-grooves, besides
  being concentrated at the nano-grooves, two energy maxima between them start to be formed, inducing
  the 1st splitting of the initially-formed nanogratings, as shown in (d). The intensity is then concentrated
  both at the newly-formed and initially-formed nano-grooves until their depth are the same, as indicated
  in (e). Due to the intensity maxima re-formed between the nano-grooves, the splitting is self-replicated
  as shown in (e)-(g).}
  \label{fig:boat1}
\end{figure}
Figure S6 demonstrates the field distributions of two periods with a series of nano-grooves’ depth. As illustrated, both at periods of 
240 nm and 120 nm, the splitting inclines to be triggered by the increase of the nano-grooves’ depth. With the pulses accumulated, the 
ridges of the nanogratings continue to be thinned by the strong intensity, through which the process of decreasing the film thickness 
and splitting influence each other until splitting halts. As a consequence, the height of the U-nanogratings was observed to remain 
constant at $\sim$40 nm (Figure 3b and 3c). The smallest period that appeared in our experiments was $\sim$35 nm. In this case, the width of the 
nano-groove was $\sim$10 nm, which is close to the extreme scale of the electrostatic regime.[53]\\
With a smaller v, more pulses are deposited on the film which leads to an increase of the surface temperature and “softer material” with lower viscosity. This allows the formation of the nanogratings with larger periods.[24] Consequently, the period of the finally formed U-nanogratings can be tuned continuously, as shown in Figure 3a. For a thicker composite film, less influence from the substrate gives rise to a larger a and thus a larger period of the C-nanogratings and U-nanogratings, which are consistent with the calculation results displayed in Figure 9 and experimental results in Figure 3b.\\
\FloatBarrier
\subsection{The comprehensive mechanism of forming the U-nanogratings}
On the basis of all the experimental results, theoretical analysis and simulation results, we propose an overall process describing the 
comprehensive formation mechanism of the U-nanogratings. In the solidified composite film, it is natural to find some defects. At the 
center of the irradiation area, the defects can become nucleation centers for nanoplasmas and develop into plasmas via multiphoton and 
avalanche ionization, causing the film to be inhomogeneous. The energy is then inclined to be concentrated into these locations in which the intensity exceeds the material modification threshold and defect-holes can be created, as shown in Figure 1b and Figure 5a. For the single spot irradiation, the sample cannot move. After the initial laser pulses ionize the area around the defects, the energy tends to be locked in the modified material, thus the interaction with the subsequent pulses will be affected. The memory mechanism and mode selection providing positive feedback can promote the growth of the initially formed nanostructures.[54] Therefore, from the periphery to the center in Figure 5a, we can observe that the defect-holes are gradually lined up into order and evolve into initial C-nanogratings at the center of the irradiation spot. The period of these HSFL is ~240 nm and in the range predicted by the SP model as shown in Figure 9.  The influence of the substrate, i.e. a varies with the depth and width of the formed C-nanogratings in the film. In addition, under different irradiation fluence, pulse number, the density of the excited electrons also changes. Thus, the period of the initially formed gratings is determined by multiple factors. From Figure 9, we can observe that the period of the gratings ranges between ~113 and ~400 nm.\\
As irradiation continued, the C-nanogratings split twice and the period “jumped” down very quickly. The period of the finally formed 
U-nanogratings can be controlled by the experimental conditions, such as the scanning speed (Figure 3a) and the film thickness (Figure 
3b). Interestingly, we did not find a clear relation between the period and the irradiation wavelength, suggesting that $\lambda$ is 
at least not the only factor to determine $\Lambda$.\\
\begin{figure}[H]
  \includegraphics[width=0.95\textwidth]{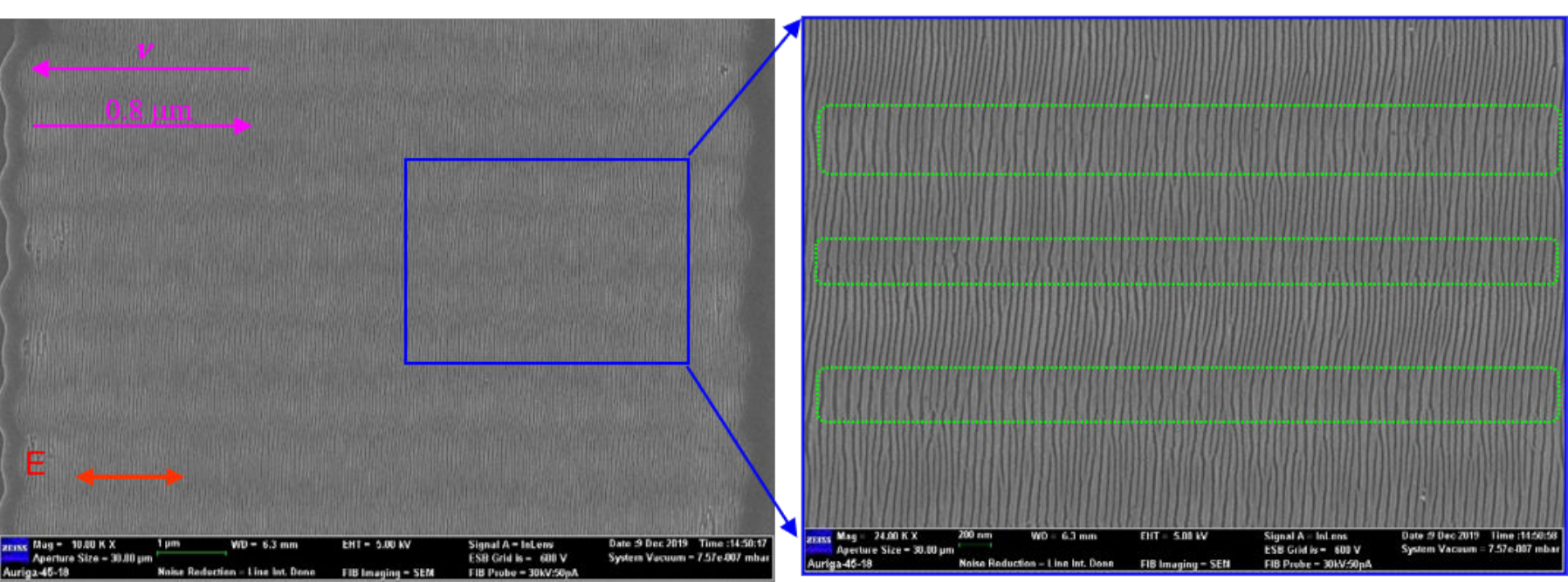}
  \centering
  \caption{A large area covered with U-nanogratings has been obtained by scanning the sample line by line. The right view zooms in the 
  details of the U-nanogratings, in which the green dashed rectangles mark the positions of the stitching areas. The fluence was $67 mJ/cm^2$ 
  and the scanning speed v=8 $\mu m/s$.}
  \label{fig:boat1}
\end{figure}
A large area covered with U-nanogratings can be obtained by multiple scanning the sample as shown in Figure 11. However, these 
U-nanogratings cannot be stitched together seamlessly and dislocations are obvious between the adjacent scanning (marked by the green 
dashed rectangles). Of course, if we apply a beam shaper to ensure a larger irradiation spot (kHz) with uniform power density, only one 
scanning can produce a large area full of regular U-nanogratings. \\

Finally, we would like to emphasize that why two sequent splitting phenomenon can be observed in our material system. Firstly, 
the introduction of Fe ions into the polymer film can reduce its Tg, which decreases its bonding energy and allows fs laser processing 
under low fluence. Therefore, even if the power density is only slightly higher on the ridge of the nanogratings, splitting can occur. 
Furthermore, a smaller Tg can enhance the segmental chain mobility in the composite PVP film,[55] which facilitates imprinting the field
 intensity distribution. On the other hand, as we have analyzed in our previous paper, Fe ions are inclined to be bound with the lone 
 pair electrons on the oxygen atom of PVP.[38] This kind of coordination bonds can act as sacrificial bonds and provide a bridge between
  the ultrashort laser pulses and the composite for ultrafast subcycle energy transfer.[56] Consequently, the hot-electron transport and
   diffusion can be ignored, i.e. the thermal effect is almost eliminated, which facilitates the second splitting and the formation of 
   the U-nanogratings.[27] \\

   \FloatBarrier
\section{Conclusions and outlook}
To summarize, we have performed a systematical investigation on the formation of ultrafine and extremely regular nanogratings with a 
period reaching 35.0 ($\pm$2.0) nm. Under intense laser irradiation, the PVP-based composite film is excited via nonlinear ionization 
processes at first. With an increase of the deposited energy and nano-grooves’ depth, two sequent grating-splitting occurs, through 
which the period is narrowed down very quickly. The unique mechanism enables the formation of U-nanogratings with periods far beyond 
that predicted by the SP model. These U-nanogratings can not only be applied to tailoring a great variety of surface properties, but 
also possibly function as masks for further etching to obtain ultrafine nanostructures efficiently. Furthermore, as the direction of 
the optical axis of the nanogratings can be controlled by the polarization direction of the irradiation beam, they are possible to 
find applications in the optical devices based on the Pancharatnam-Berry phase.[57] Ultimately, the proposed formation mechanism 
encourages us to incorporate other metallic ions into the PVP polymer and irradiate them with a laser fluence reaching their ablation 
threshold to generate various metallic ultrafine nanostructures with diverse functionalities.\\

\FloatBarrier
\medskip
\textbf{Supporting Information} \par 
Supporting Information is available from the Wiley Online Library or from the author.

\medskip
\textbf{Acknowledgements} \par 
The authors acknowledge Prof. Sheng Lan and Dr. Haihua Fan for their support on the optical
equipment devices and Prof. Juntao Li on the SEM imaging. SEM was provided by the State
Key Laboratory of Optoelectric Materials and Technologies in Sun Yat-sen University. This
work was funded by National Natural Science Foundation of China (NSFC) (Grant Nos.61675070 and 11704133) 
and Science and Technology Program of Guangzhou (No.2019050001)

\medskip

%

\medskip
\textbf{Author Contributions} \par 
The manuscript was written through contributions of all authors. All authors have given
approval to the final version of the manuscript. ‡These authors contributed equally.
\medskip

\textbf{References}\\

[1]	P. Meyrueis, D. V. M. Van, K. Sakoda, Micro- and Nanophotonic Technologies 2017, Wiley-VCH, Germany, 2017.\\{}
[2]	P. N. Prasad, Introduction to Biophotonics, Wiley-Interscience, New York, 2003.\\{}
[3]	T. J. Palinski, B. E. Vyhnalek, G. W. Hunter, A. Tadimety, J. X. J. Zhang, IEEE J. Sel. Top. Quantum Electron. 2021, 27, 1.\\{}
[4]	K. Sangeeth, G. M. Hegde, Curr. Sci. 2014, 107, 749.\\{}
[5]	C. J. Bettinger, R. Langer, J. T. Borenstein, Angew. Chem. Int. Ed.. 2009, 48, 5406.\\{}
[6]	B. Bhushan, Encyclopedia of Nanotechnology. Springer Netherlands,2016.\\{}
[7]	X. Zhang, M. Theuring, Q. Song, W. Mao, M. Begliarbekov, S. Strauf, Nano Lett. 2011, 11, 2715.\\{}
[8]	J. Perriere, E. Millon, E. Fogarassy, Recent Advances in Laser Processing of Materials, Elsevier Science Ltd, Amsterdam, 2006.\\{}
[9]	G. J. Lee, S. H. Song, Y. P. Lee, H. Cheong, J. Jin, Appl. Phys. Lett. 2006, 89, 151907.\\{}
[10]	D. Tan, Y. Li, F. Qi, H. Yang, Q. Gong, X. Dong, X. Duan, Appl. Phys. Lett. 2007, 90, 71106.\\{}
[11]	Mojarad,N.,Hojeij,M.,Wang,L.,Gobrecht,J.,Ekinci,Y., Nanoscale. 2013, 7, 4031.\\{}
[12]  Z. Gan, Y. Cao, R. A. Evans, M. Gu, Nat. Commun. 2013, 4, 2061.\\{}
[13]	M. Birnbaum, J. Appl. Phys. 1965, 36, 3688.\\{}
[14]	J. S. Preston, H. M. V. Driel, J. E. Sipe, Phys. Rev. B.1989,40, 3942.\\{}
[15]	J. E. Sipe, J. F. Young, J. S. Preston, H. M. Van Driel, Phys. Rev. B. 1983, 27, 1141.\\{}
[16]	A. Rudenko, J. Colombier, S. Höhm, A. Rosenfeld, J. Krüger, J. Bonse, T. E. Itina, Sci. Rep. 2017, 7, 12306.\\{}
[17]	J. Bonse, A. Rosenfeld, J. Krüger, J. Appl. Phys. 2009, 106, 104910.\\{}
[18]	M. Huang, F. Zhao, Y. Cheng, N. Xu, Z. Xu, ACS Nano 2009, 3, 4062.\\{}
[19]	R. Buividas, M. Mikutis, S. Juodkazis, Prog. Quantum Electron. 2014, 38,119.\\{}
[20]	J. Bonse, S. Gräf, Laser Photon. Rev. 2020, 14, 2000215.\\{}
[21]	J. Bonse, S. Hohm, S. V. Kirner, A. Rosenfeld, J. Kruger, IEEE J. Sel. Top. Quantum Electron. 2017, 23, 9000615.\\{}
[22]	M. Huang, F. Zhao, Y. Cheng, N. Xu, Z. Xu, Phys. Rev. B, Condens. Matter. 2009, 79, 125436.\\{}
[23]	E. Rebollar, J. R. Vázquez De Aldana, J. A. Pérez-Hernández, T. A. Ezquerra, P. Moreno, M. Castillejo, Appl. Phys. Lett. 2012, 100, 41106.\\{}
[24]	E. Rebollar, M. Castillejo, T. A. Ezquerra, Eur. Polym. J. 2015, 73,162.\\{}
[25]	X. He, A. Datta, W. Nam, L. M. Traverso, X. Xu, Sci. Rep. 2016, 6, 35035.\\{}
[26]	M. Mezera, M. V. Drongelen, G. Rmer, J. Laser Micro Nanoeng. 2018,13,105.\\{}
[27]	M. Huang, Y. Cheng, F. Zhao, Z. Xu, Ann. Phys. 2013, 525, 74.\\{}
[28]	E. Blasco, J. Müller, P. Müller, V. Trouillet, M. Schn, T. Scherer, C. B. Kowollik, M. Wegener, Adv. Mater. 2016, 28, 3592.\\{}
[29]	F. Kotz, K. Arnold, W. Bauer, D. Schild, N. Keller, K. Sachsenheimer, T. M. Nargang, C. Richter, D. Helmer, Rapp, B. E. Nature 2017, 544, 337.\\{}
[30]	P. Franco, M. I. De, Polymers 2020, 12, 1114.\\{}
[31]	M. Teodorescu, M. Bercea, S. Morariu, Biotechnol. Adv. 2019, 37, 109.\\{}
[32]	L. D. Zarzar, B. S. Swartzentruber, J. C. Harper, D. R. Dunphy, C. J. Brinker, J. Aizenberg, B. Kaehr, J. Am. Chem. Soc. 2012, 134, 4007.\\{}
[33]	K. Vora, S. Kang, S. Shukla, E. Mazur, Appl. Phys. Lett. 2012, 100, 63120.\\{}
[34]	S. Shukla, X. Vidal, E. P. Furlani, M. T. Swihart, K. Kim, Y. Yoon, A. Urbas, P. N. Prasad, ACS Nano 2011, 5, 1947.\\{}
[35]	M. Askari, D. A. Hutchins, P. Thomas, L. Astolfi, A. T. Clare, Addit. Manuf. 2020, 36,101562.\\{}
[36]	M. Jin, X. Zhang, S. Nishimoto, Z. Liu, D. A. Tryk, T. Murakami, A. Fujishima, Nanotechnology. 2007, 18, 075605.\\{}
[37]	L. C. Lopérgolo, A. B. Lugão, L. H. Catalani, Polymer 2003, 44, 6217.\\{}
[38]	M. Pan, C. Guo, Y. Wang, Z. Ma, L. Chen, Q. Li, L. Wu, Chem. Phys. Lett .2020, 754, 137640.\\{}
[39]	Y. Lei, M. Sakakura, L. Wang, Y. Yu, H. Wang, G. Shayeganrad and P. G. Kazansky, Optica 2021, 8,1365.\\{}
[40]	M. Huang, F. Zhao, Y. Cheng, N. Xu, Z. Xu, Opt. Express. 2008, 15, 19354.\\{}
[41]	C. Yao, Y. Ye, B. Jia, Y. Li, R. Ding, Y. Jiang, Y. Wang, X. Yuan, Appl. Surf. Sci. 2017, 425, 1118.\\{}
[42]	F. Zimmermann, A. Plech, S. Richter, A. Tünnermann and S. Nolte, Laser Photon. Rev. 2016, 10, 327.\\{}
[43]	J. W. Yao, C. Y. Zhang, H. Y. Liu, Q. F. Dai, L. J. Wu, S. Lan, A. V. Gopal, V. A. Trofimov, and T.M. Lysak, Opt. Express 2012, 20, 905.\\{}
[44]	Y. Liao, W. Pan, Y. Cui, L. Qiao, Y. Bellouard, K. Sugioka, Y. Cheng, Opt. Lett. 2015, 40, 3623.\\{}
[45]	E. Rebollar, S. Pérez, J. J. Hernández, I. Martín-Fabiani, D. R. Rueda, T. A. Ezquerra, M. Castillejo, Langmuir 2011, 27, 5596.\\{}
[46]	B. Chimier, O. Utéza, N. Sanner, M. Sentis, T. Itina, P. Lassonde, F. Légaré, F. Vidal, J. C. Kieffer, Phys. Rev. B 2011, 84, 094104.\\{}
[47]	M. Malinauskas, A. Zukauskas, G. Bickauskaite, R. Gadonas, S. Juodkazis, Opt. Express 2010, 18, 10209.\\{}
[48]	S. K. Das, H. Messaoudi, A. Debroy, E. McGlynn, R. Grunwald, Opt. Mater. Express 2013, 3, 1705.\\{}
[49]	A. Dereux, T. W. Ebbesen, W. L. Barnes, Nature 2003, 424, 824.\\{}
[50]	L. Wang, B. Xu, X. Cao, Q. Li, W. Tian, Q. Chen, S. Juodkazis, H. Sun, Optica 2017, 4, 637.\\{}
[51]	T. J. Y. Derrien, R. Koter, J. Krüger, S. Höhm, A. Rosenfeld, J. Bonse, J. Appl. Phys. 2014, 116, 074902.\\{}
[52]	L. Wang, X. Cao, C. Lv, H. Xia, W. Tian, Q. Chen, S. Juodkazis, H. Sun, IEEE J. Quantum Electron. 2018, 54, 9200207.\\{}
[53]	M. Huang, Z. Xu, Laser Photon. Rev. 2014, 8, 633.\\{}
[54]	V. R. Bhardwaj, E. Simova, P. P. Rajeev, C. Hnatovsky, R. S. Taylor, D. M. Rayner, P. B. Corkum, Phys. Rev. Lett. 2006, 96, 057404.\\{}
[55]	B. Zhang, K. Kowsari, A. Serjouei, M. L. Dunn, Q. Ge, Nat. Commun. 2018, 9, 1831.\\{}
[56]	Y. Liu, W. Cao, M. Ma, P. Wan, ACS Appl. Mater. Inter. 2017, 9, 25559.\\{}
[57]	E. Hasman, V. Kleiner, G. Biener, A. Niv, Appl. Phys. Lett. 2003, 82, 328.\\{}

\end{document}


\pagestyle{fancy}

\title{
  \centerline{\textbf{Supporting Information}}
}

\maketitle
\title{Sub-40nm Nanogratings Self-Organized in PVP-based Polymer Composite Film by Photoexcitation and Two Sequent Splitting under Femtosecond Laser Irradiation}

\maketitle


\author{Li-Yun Chen$\ddagger $,}
\author{Cheng-Cheng Guo$\ddagger $,}
\author{Ming-Ming Pan,}
\author{Chen Lai,}
\author{Yun-Xia Wang,}
\author{Guo-Cai Liao,}
\author{Zi-Wei Ma,}
\author{Fan-Wei Zhang,}
\author{Jagadeesh Suriyaprakash,}
\author{Lijing Guo,}
\author{Eser Akinoglu,}
\author{Qiang Li$^*$,}
\author{ Li-Jun Wu$^*$}

\begin{affiliations}
  L.-J. Wu, Q. Li, L.-Y. Chen, C.-C. Guo, M.-M. Pan, C. Lai, Y.-X. Wang, G.-C. Liao, Z.-W. Ma, F.-W. Zhang, 
  J. Suriyaprakash\\
  Guangdong Provincial Key Laboratory of Nanophotonic Functional Materials and Devices School of Information and Optoelectronic Science and Engineering\\
  South China Normal University\\ 
  Guangzhou 510006, China\\
  
  Email Address:ljwu@scnu.edu.cn; liqiangnano@m.scnu.edu.cn\\
 
  \medskip
  L.J. Guo, E. Akinoglu\\
  International Academy of Optoelectronics at Zhaoqing\\
  South China Normal University\\
  Guangzhou 510006, China\\
\end{affiliations}


\section{Calculation of the threshold of forming the U-nanogratings}
For a Guassian spatial profile, the fluence F(z,r) is given by the following equation, [1]\\

\begin{equation}
  F(z,r)=F_{0} \frac{w_{0}^{2} }{w^{2}(z) } exp(-2r^{2}/w^{2}(z)  )
\end{equation}
where z is the focus position and r is the radius. F$_0$ is the peak fluence at r=0 and z=0.  w0 represents the $1/e^2$ beam waist 
at the focus position (z=0). The beam waist w (z) is dependent on z and the wavelength $\lambda $: \\
\begin{equation}
  w(z)=w_{0} \sqrt{1+\frac{z^{2}\lambda ^{2}  }{\pi ^{2}w_{0}^{4}  } }  
\end{equation}
The correlation to the pulse energy E$_{pulse}$ is given by the integral over the profile: \\
\begin{equation}
  E_{p u l s e}=F_{0} \frac{w_{0}^{2}}{w^{2}(z)} \int_{0}^{\infty} \exp \left(-2 r^{2} / w^{2}(z)\right) \times 2 \pi r d r=\frac{1}{2} \pi F_{0} w_{0}^{2}
\end{equation}
From Equation (3), we can see that Epulse is not related with the z-position. When z=0, \\
\begin{equation}
  E_{pulse}=\frac{1}{2}  \pi F_{0} w_{0}^{2} = \pi F w_{0}^{2}
\end{equation}
\begin{equation}
  F=\frac{E_{pulse} }{\pi w_{0}^{2} } =\frac{1}{2} F_{0} 
\end{equation}
F is the average fluence. The diameter of the focused beam is $\sim$ 1 $\mu m$ in our irradiation set up (with a repetition rate of 76 MHz) 
and the fluence can be considered as a Gaussian profile in a 2 $\mu m$ area. Combined with the SEM result, we can then calculate the 
threshold ($F^*$) of forming the U-nanogratings as shown in Figure 2a. \\

\begin{figure}[p]
  \includegraphics[width=0.95\textwidth]{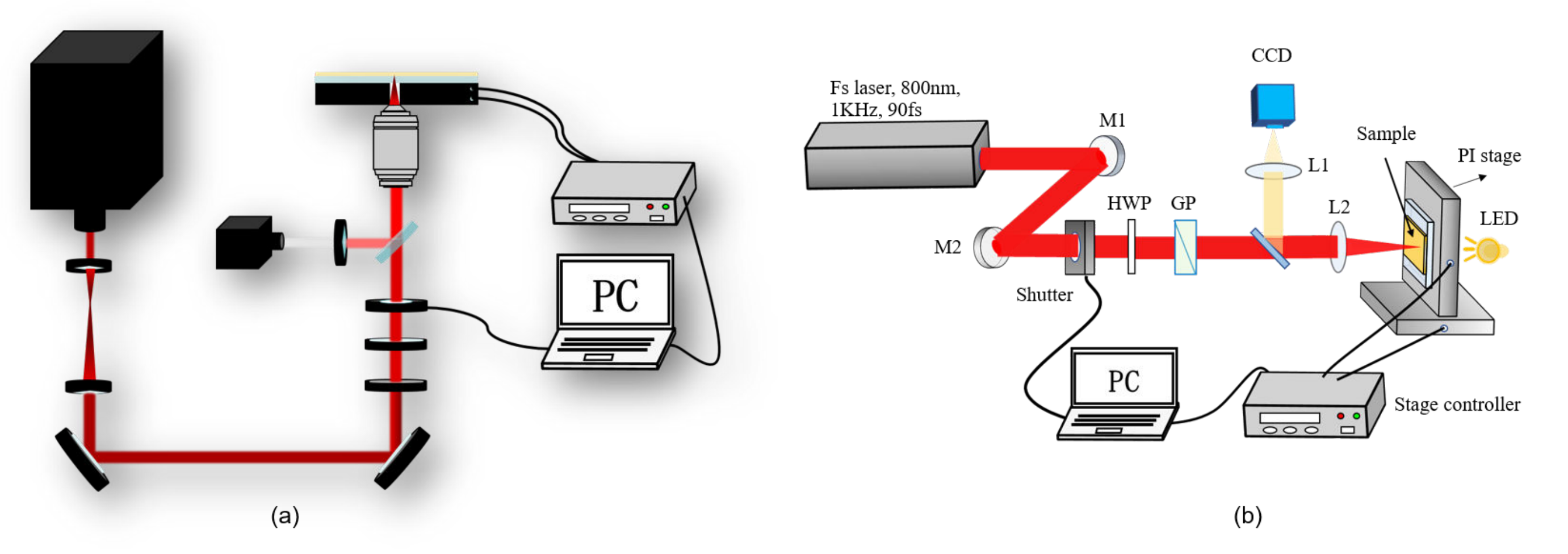}
  \caption{The schematics of the irradiation set up with the laser repetition rate of (a) 76 MHz and (b) 1 KHz. 
  HWP: half wave-plate; GP: Glan polarizer; DI: dichroscope M1 and M2 are two mirrors. L1 and L2 are two lens. }
  \label{fig:boat1}
\end{figure}

\begin{figure}[p]
  \includegraphics[width=0.5\textwidth]{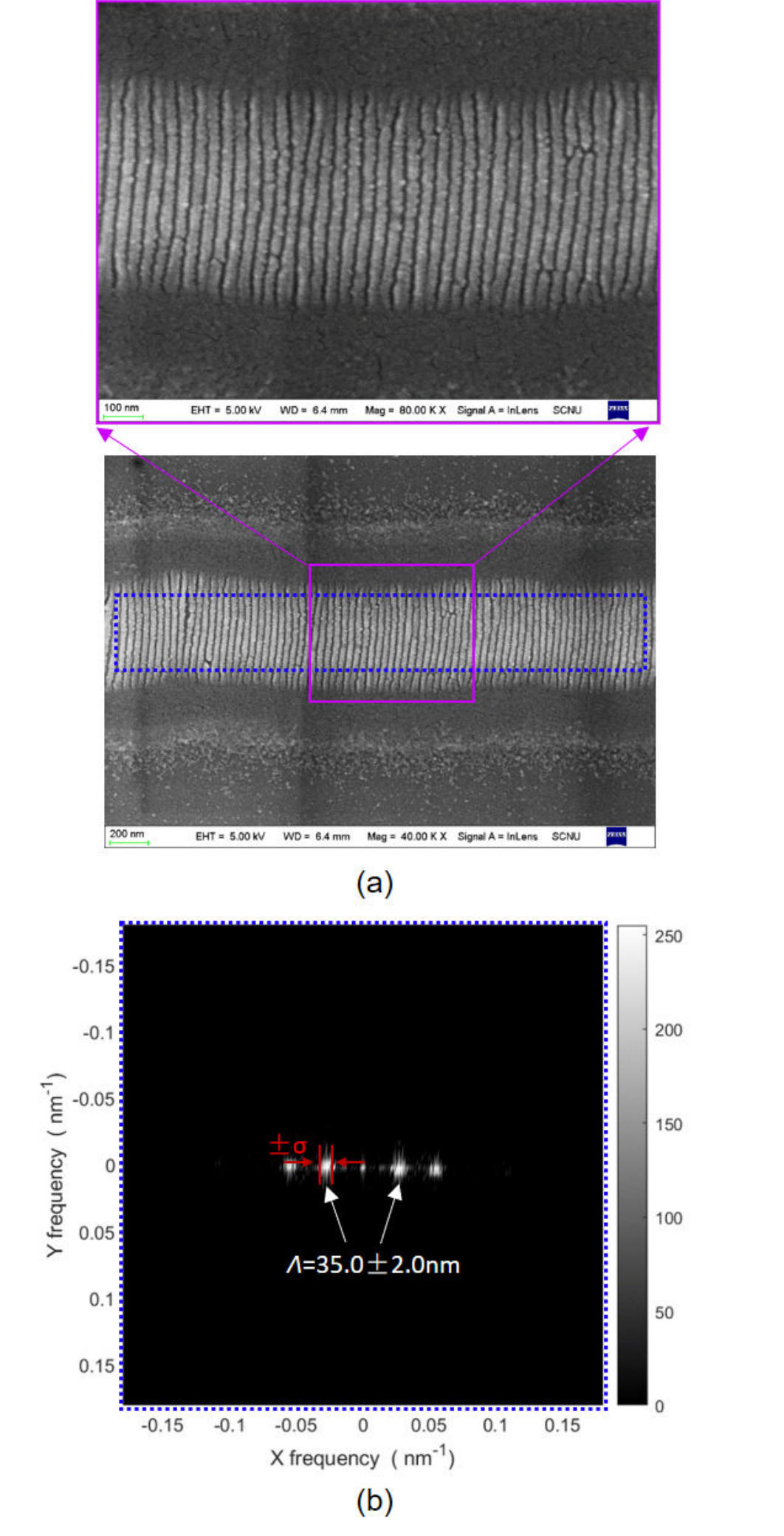}
  \centering
  \caption{(a) A typical SEM image of the lines written by scanning the sample with a speed of 16 $\mu m$/s when the fluence 
  was 25 $mJ/cm^2$ at a wavelength of 532 nm. The top panel represents a zoomed-in portion. (b) The 2D Fourier Transform (2D-FT)  
  result for the nanogratings in the blue dashed square. The period of the nanogratings is measured to be $\Lambda $=35.0$\pm $2.0 nm. }
  \label{fig:boat1}
\end{figure}

\begin{figure}[p]
  \includegraphics[width=0.95\textwidth]{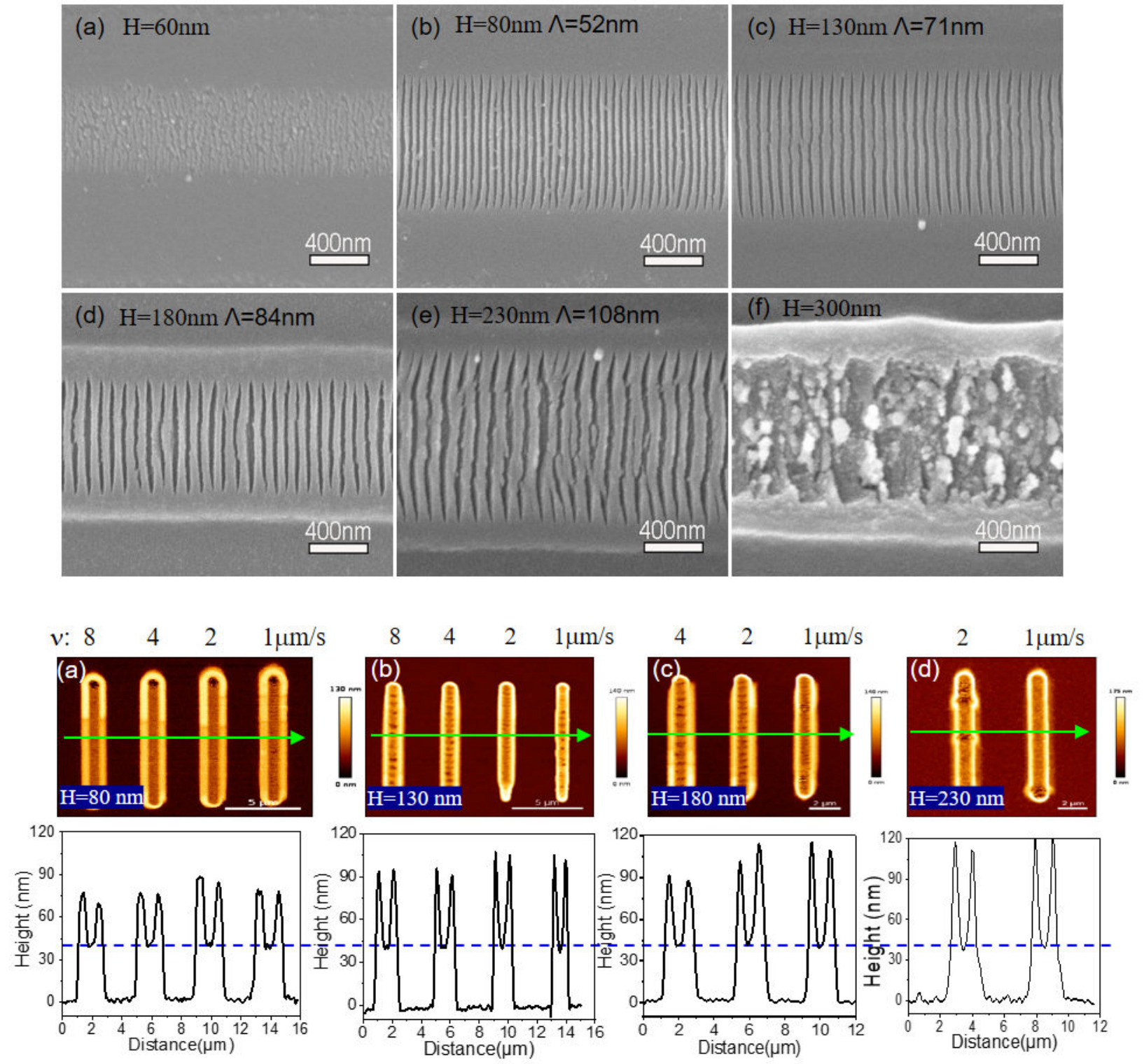}
  \caption{Top panel: SEM results for the laser scanned lines under different thickness of the composite film. The measured periods 
  are shown in the insets. The irradiation fluence F=42 $mJ/cm^2$ and scanning speed $\upsilon$ =1 $\mu m/s$. Bottom panel: Corresponding AFM results.  }
  \label{fig:boat1}
\end{figure}

\begin{figure}[p]
  \includegraphics[width=0.95\textwidth]{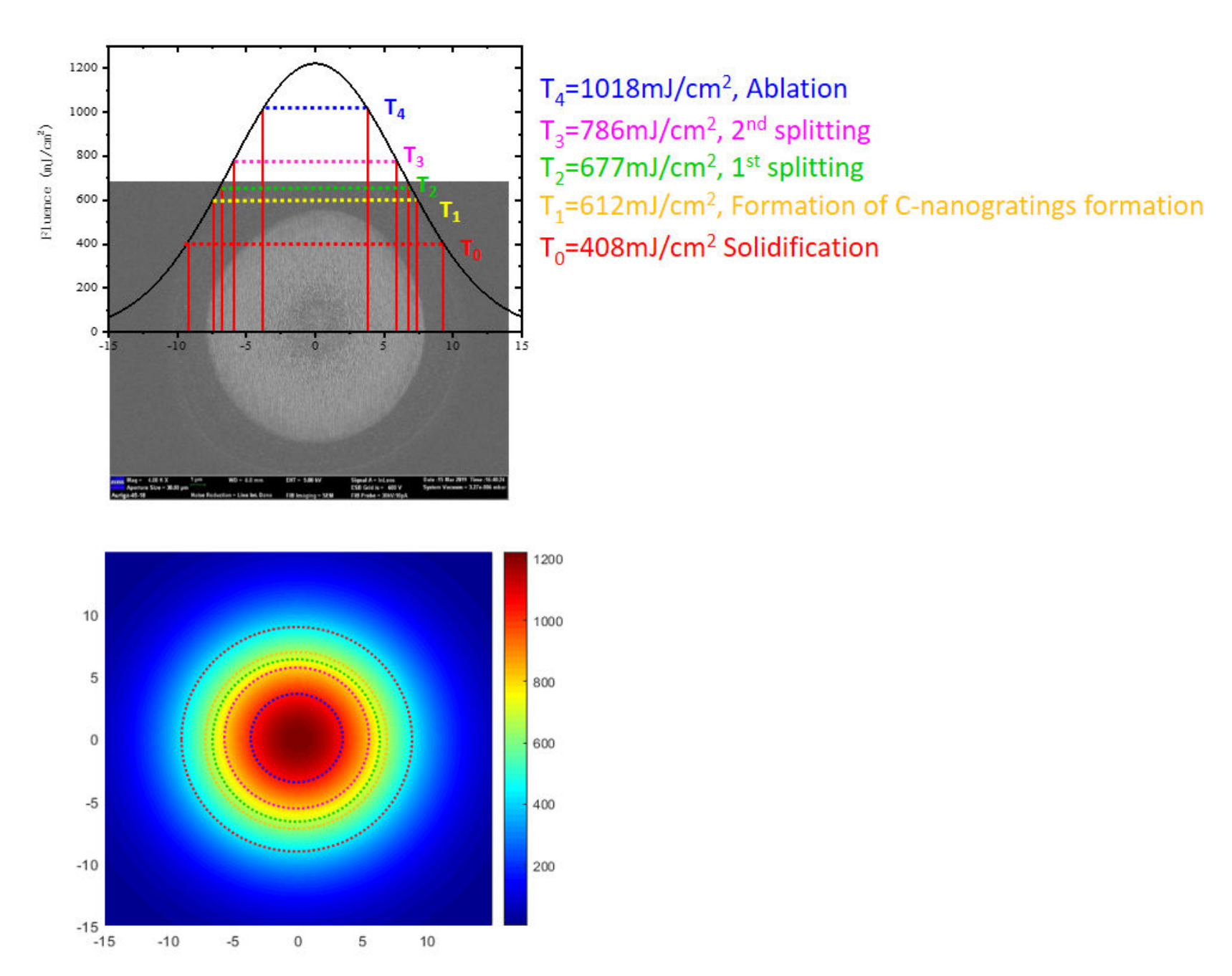}
  \caption{Calculation of the threshold for solidification, formation of C-nanogratings, 1st splitting, 2nd splitting and ablation 
  with a fluence of 600 $mJ/cm^2$ (repetition rate 1 kHz) based on a Gaussian spatial profile and the SEM result. }
  \centering
  \label{fig:boat1}
\end{figure}

\FloatBarrier
\section{Calculation of the dielectric constant $\varepsilon ^*$ of the excited composite film and laser excited electron concentration Ne }
The permittivity of the excited composite film can be described by the Drude equation [2,3] and its dielectric constant $\varepsilon ^*$ varies 
with the laser excited electron concentration $N_e$.[4, 5, 6]\\
\begin{equation}
  \varepsilon^{*}=1+\left(\varepsilon_{c}-1\right)\left(1-\frac{N_{e}}{N_{v}}\right)-\frac{\omega_{p}^{2}}{\omega^{2}} \frac{1}{\left(1+i\left(\omega \tau_{c}\right)^{-1}\right)}
\end{equation}

\begin{equation}
  \omega_{\mathrm{p}}^{2}=\frac{\mathrm{e}^{2} \mathrm{~N}_{\mathrm{e}}}{\mathrm{m}^{*} \varepsilon_{0}}
\end{equation}
$\varepsilon_c $is the static dielectric constant of the composite film and was measured to be 2.45. $\varepsilon _o$ is the dielectric constant of vacuum. 
$\omega $ and $\tau _{c} $ is the frequency of the irradiation light and the electron damping time respectively. $\omega _p $is the plasmonic frequency and 
related with Ne and the effective electron mass $m^*$. Nv is the initial density of the electrons in the valence band and can be
calculated by,[7]\\

\begin{equation}
  N_{v}=X_{0} \rho N_{a} / M
\end{equation}
where Na is the Avogadro constant. The density $\rho $ and the monomer molar mass M of the composite film can be approximated to be 1200 $kg/m^3$ 
and 111 g/mol respectively.[8] $X_{0}$ is the total valence number of electrons in the monomer and equals to $4\times 6+1\times9+3+2=38$ in each molecular 
of PVP with formula $C_6H_9NO$. Nv is then calculated to be $2.47\times10^{23}$ $cm^{-3}$. When the real part of the effective dielectric constant, 
Re[$\varepsilon^*$]=0, the material is normally considered to be optically broken down and turns from a dielectric to a metallic state.[5] 
In this case, Ne is calculated to be $4\times10^{21} cm^{-3}$.\\
Based on Eq. (6)-(8), $\varepsilon^*$ can be calculated out. Actually, it is hard to obtain the precise value of the parameters in the composite 
film and some of them such as $\tau c$ and $m^*$ have to be estimated. The influence of $\tau_{c}$ and $m^*$  on Ne is described in the process of 
avalanche and multiphoton ionization respectively,[9, 10, 11] which are plotted in Figure S5. As shown, both $\tau_{c}$ and $m^*$ do not have 
obvious effect on Ne. Without loss of generality, we set $\tau_{c}$ =2.5 fs and $m^*$ =0.45$m_{e}$ by referring to the typical values applied in other
 polymers.[7]\\

\begin{figure}[p]
  \includegraphics[width=0.95\textwidth]{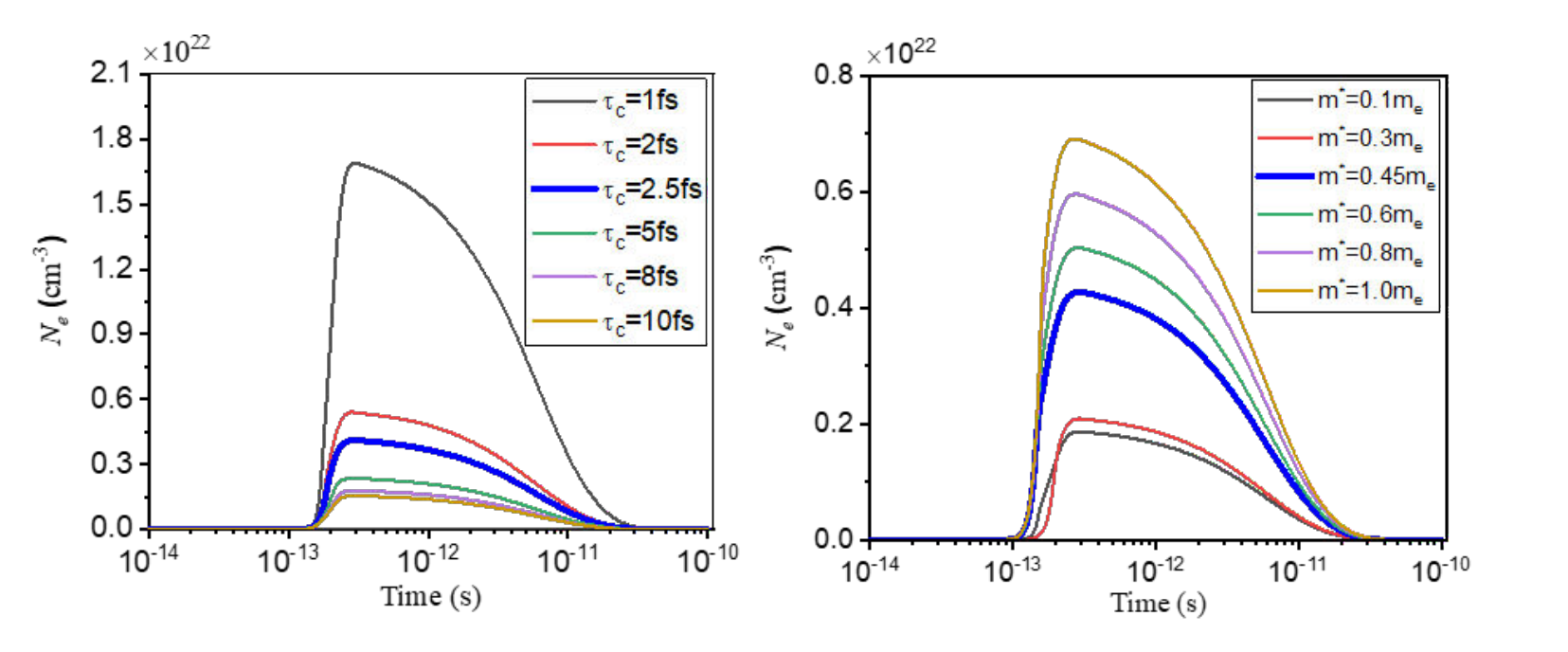}
  \caption{The change of the excited carrier density Ne upon fs laser irradiation under (a) different electron damping time $\tau_{c}$ 
  with $m^*$ =0.45$m_{e}$ and (b) different effective electron mass  $m^*$ with $\tau_{c}$ =2.5 fs. The thicker blue lines represent the parameters 
  applied in the paper }
  \label{fig:boat1}
\end{figure}

\begin{figure}[htp]
  \includegraphics[width=0.95\textwidth]{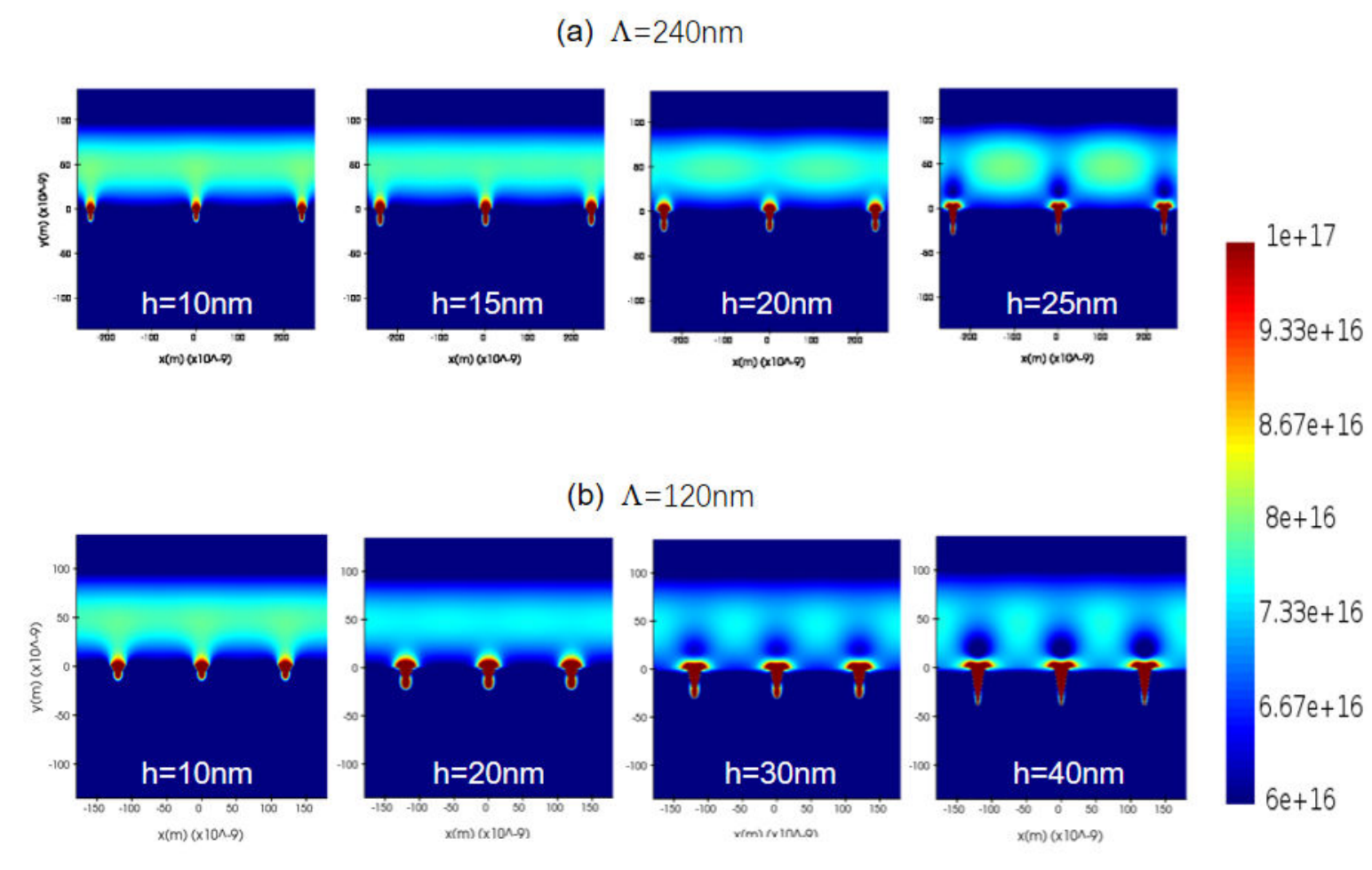}
  \caption{The distributions of the light intensity with a period $\Lambda $ of (a) 240 nm and (b) 120 nm under different nano-grooves’ depth. }
  \label{fig:boat1}
\end{figure}

\FloatBarrier
\medskip
\textbf{References}\\

[1]	Gregorcic P.,Sedlacek M.,Podgornik B.,Reif J.,Applied Surface Science 2016, 387, 698.\\{}
[2]	J. Bonse, S. Gräf, Laser Photon. Rev. 2020, 14, 2000215.\\{}
[3]	L. Wang, B. Xu, X. Cao, Q. Li, W. Tian, Q. Chen, S. Juodkazis, H. Sun, Optica 2017, 4, 637.\\{}
[4]	J. Bonse, S. Hohm, S. V. Kirner, A. Rosenfeld, J. Kruger, IEEE J. Sel. Top. Quantum Electron. 2017, 23, 9000615.\\{}
[5]	Y. Liao, W. Pan, Y. Cui, L. Qiao, Y. Bellouard, K. Sugioka, Y. Cheng, Opt. Lett. 2015, 40, 3623.\\{}
[6]	T. J. Y. Derrien, R. Koter, J. Krüger, S. Höhm, A. Rosenfeld, J. Bonse, J. Appl. Phys. 2014, 116, 074902.\\{}
[7]	L. Wang, X. Cao, C. Lv, H. Xia, W. Tian, Q. Chen, S. Juodkazis, H. Sun, IEEE J. Quantum Electron. 2018, 54, 9200207.\\{}
[8]	S. Pérez, E. Rebollar, M. Oujja, M. Martín, M. Castillejo, Appl. Phys. A. 2013, 110, 683.\\{}
[9]	M. Malinauskas, A. Zukauskas, G. Bickauskaite, R. Gadonas, S. Juodkazis, Opt. Express 2010, 18, 10209.\\{}
[10]	B. C. Stuart, M. D. Feit, S. Herman, A. M. Rubenchik, B. W. Shore, M. D. Perry, Phys. Rev. B. 1996, 53, 1749.\\{}
[11]	G. D. Tsibidis, E. Skoulas, A. Papadopoulos, E. Stratakis, Phys. Rev. B 2016, 94,081305.\\{}